\documentclass[aps,prd,preprintnumbers,nofootinbib,floatfix,11pt,a4paper]{revtex4-2}

\usepackage{amsfonts,amsmath,amssymb,mathrsfs,graphicx,epsfig,color,amsthm} 
\usepackage{hyperref}
\hypersetup{
  colorlinks=true,
  linkcolor=blue,
  citecolor=blue,
  urlcolor=blue,
  bookmarksnumbered=true
}

\allowdisplaybreaks[1]
\newcommand{\be}{\begin{equation}}
\newcommand{\ee}{\end{equation}}
\newcommand{\ba}{\begin{eqnarray}}
\newcommand{\ea}{\end{eqnarray}}

\theoremstyle{definition}

\usepackage[normalem]{ulem}

\begin{document}

\title{A simple flux criterion for trapped surface formation
by short pulses of gravitational radiation and matter}
\author{Tetsuya Shiromizu$^{1,2,3}$, Keisuke Izumi$^{1,2}$ and Daisuke Yoshida$^3$
}

\affiliation{$^{1}$Department of Mathematics, Nagoya University, Nagoya 464-8602, Japan}
\affiliation{$^{2}$Kobayashi-Maskawa Institute, Nagoya University, Nagoya 464-8602, Japan} 
\affiliation{$^{3}$Institute for Advanced Research, Nagoya University, Nagoya 464-8602, Japan} 

\begin{abstract}
\noindent

We discuss the formation of trapped surfaces induced by short pulses of incoming gravitational radiation and matter fields, 
in a double-null formulation.
We identify a hierarchy of orders in the pulse width for the geometric and matter quantities consistent with the evolution equations. 
Under the assumption that this hierarchy holds throughout the region considered,
the incoming fluxes undergo approximately conservative 
inward transport and are thereby amplified. 
The Raychaudhuri equation then gives an initial flux criterion for trapped surface formation.
The conditions on the matter are expressed through the Ricci components and their derivatives, 
which allows the argument to be developed without specifying a particular matter model.

\end{abstract}

\maketitle


\section{Introduction}

The formation of trapped surfaces is one of the central geometric mechanisms underlying black-hole formation in general relativity \cite{Penrose1965,HawkingEllis}. 
A trapped surface is a closed spacelike two-surface for which both future-directed null expansions are negative. 
A remarkable development in this direction was Christodoulou's proof of trapped surface formation through the strong focusing of 
null geodesic congruences by short pulses of incoming gravitational waves \cite{Christodoulou2009}. This result established that trapped surfaces
can form dynamically in vacuum, without assuming
any spacetime symmetries. 
This provides rigorous mathematical support for the physical picture suggested by earlier simulations by Miyama and, subsequently, 
by Abrahams and Evans, which indicated that sufficiently strong and spatially concentrated gravitational-wave packets can collapse 
under their own gravity and form apparent horizons~\cite{Miyama1981,AbrahamsEvans1992}.

Subsequent mathematical studies \cite{KlainermanRodnianski2012,AnNullInfinity,KlainermanLukRodnianski2013,AnLuk2017,An:2019ybm}
have refined and extended this short-pulse mechanism, giving a mathematically precise understanding of how sufficiently concentrated 
incoming radiation can lead to trapped surface formation.
Related results have also been established for several Einstein--matter systems, with sufficient conditions involving the combined incoming flux 
of gravitational radiation and matter~\cite{Yu:2011qy,AnAthanasiou2020,ZhaoHilditchValienteKroon2023,AthanasiouMondalYau2023,Shen:2025ztd}.
For the Einstein--Maxwell system, it has been shown that the concentration of 
the electromagnetic field alone, with no incoming gravitational-wave flux, 
already leads to trapped surface formation~\cite{Yu:2011qy,AnAthanasiou2020}. 
In each case, however, the analysis relies on the field equations of the 
specific matter model under consideration.

We organize the short-pulse problem into three closely related tasks\footnote{
In Refs.~\cite{Christodoulou2009,KlainermanRodnianski2012} these steps are not 
explicitly separated: the consistency of the orders is built into the norm 
estimates. We separate them here because the present analysis is pointwise.
}: 
(i) identifying a hierarchy of orders in the pulse width and checking that it 
is consistent with the evolution equations;
(ii) establishing these estimates in suitable norms, which ensures sufficient regularity 
of the solution and its existence throughout the region;
(iii) deriving a sufficient condition for trapped surface formation under these 
estimates. 
A related distinction, between establishing the required estimates and deriving a formation criterion under those estimates, is illustrated in Secs.~1.3 and 1.4 of 
Ref.~\cite{KlainermanRodnianski2012}. 
A pedagogical account of the formation argument is given in Sec.~2 of 
Ref.~\cite{BiniEsposito2017}. 
In the present paper, we address tasks (i) and (iii), assuming the outcome of task (ii), 
{\it i.e.}, that a regular spacetime solution exists throughout the 
region considered and that the required estimates hold there. 
In vacuum, this has been established in 
Refs.~\cite{Christodoulou2009,KlainermanRodnianski2012}.

We first describe the focusing mechanism, which is task (iii).
As the incoming pulse propagates inward, the cross-sectional area of the 
incoming null congruence decreases. 
Under suitable conditions, the gravitational-wave and matter fluxes 
undergo approximately conservative transport and are therefore amplified as 
this area shrinks. 
These amplified fluxes reduce the expansion of outgoing null geodesics, as 
described by the Raychaudhuri equation. 
If the accumulated focusing effect is large enough, the outgoing null 
expansion, which is initially positive, becomes negative on a later 
two-surface. 
Since the incoming null expansion is negative there, 
this later surface becomes trapped. 
The contribution of the shear of the null congruence is interpreted as the 
energy flux of gravitational radiation, while the Ricci term represents the 
matter flux. 
This clarifies that, in non-vacuum cases, both gravitational radiation and 
matter enter on an equal footing through their energy fluxes. 
The criterion is expressed at each angle in terms of the initial flux of 
gravitational radiation and matter integrated over the pulse width.

The focusing mechanism described above relies on the approximately 
conservative transport of the two fluxes. 
We now turn to task (i), which determines the orders that such transport 
requires.
In vacuum, the required control amounts to preventing the energy of the 
infalling gravitational waves from leaking into other directions; the 
corresponding estimates are established in Refs.~\cite{Christodoulou2009,KlainermanRodnianski2012}. 
When matter is present, two further effects must be controlled: the 
scattering of gravitational-wave energy through its interaction with matter, 
and the scattering of the infalling matter itself out of the incoming null 
direction. 
The evolution equations for the two fluxes can be arranged in a similar form, and the 
contracted Bianchi identity identifies the terms that modify conservative 
transport of the matter flux. 
Requiring these terms to be subleading gives conditions on the Ricci tensor 
and its derivatives, which allow the formation argument to be developed 
without specifying a particular matter model. 
In the vacuum limit, the geometric orders assumed here coincide with the 
pointwise bootstrap bounds of Ref.~\cite{KlainermanRodnianski2012}.

We consider a double-null setting in which a regular causal diamond is 
followed by a short incoming pulse of gravitational radiation and matter. 
The central causal diamond is taken to be a region of Minkowski spacetime, 
while non-trivial incoming flux is prescribed as initial data on a null 
hypersurface extending outward from the spatial edge of this region. 
We keep the radius of the final two-surface fixed as the pulse width is made 
small.

The paper is organized as follows. In Sec.~\ref{basics}, we introduce the double-null formalism and collect the basic geometrical equations for 
the null expansions, shear, connection terms, and Ricci terms, with the Ricci terms interpreted as matter contributions through the Einstein equation.
In Sec.~\ref{TSF}, we apply these equations to a short-pulse configuration and 
derive the flux criterion for trapped surface formation. We also discuss how the flux on the final null hypersurface 
where a trapped surface is expected to form can be related to 
the initial flux. 
In Sec.~\ref{sec:order-consistency}, we introduce the hierarchy of orders, derive 
from it the bounds assumed in Sec.~\ref{TSF}, and examine its consistency 
with the evolution equations.
In Sec.~\ref{sec:summary}, we summarize and discuss our results. Appendix~\ref{app:connection} gives the evolution equations
for the connection terms,
Appendix~\ref{app:angular-commutators} gives the commutator
estimates used to extend the consistency checks to angular derivatives,
and Appendix~\ref{app:curvature} gives the curvature estimates.
Appendices~\ref{Sec+Dv+} and \ref{SecD2t} discuss the
stronger angular estimates used to transport the
gravitational-wave energy flux from the initial null
hypersurface to the final one. We follow Wald's abstract
index notation \cite{Wald1984}.

\section{Double-null formalism and basic equations}
\label{basics}

We consider a spacetime region equipped with a double-null foliation defined by two families of null hypersurfaces labeled by $u^+$ and $u^-$.
Their intersections are spacelike two-surfaces, on which we introduce local coordinates $x^A$ $(A=2,3)$.
We assume that this foliation is regular throughout the region considered.
In the double-null coordinates $\lbrace u^+,~u^-,~x^A \rbrace$, we write the metric as
\begin{equation}
g=-4\Omega^2du^+du^-+\gamma=-4\Omega^2du^+du^-+\gamma_{AB}(dx^A-\beta^A du^-)(dx^B-\beta^Bdu^-).
\end{equation}
Here, $\gamma$ is the induced metric on the spacelike two-surfaces, $\gamma_{AB}$ are its components in the coordinates $x^A$, $\Omega$ is a positive function, and $\beta^A$ is the shift vector.
We define the vector fields normal to the null hypersurfaces $u^-=\mathrm{const.}$ and $u^+=\mathrm{const.}$, respectively, by
\begin{equation}
k_+^a=-2\nabla^a u^-=\Omega^{-2}(\partial_+)^a \label{k+}
\end{equation}
and 
\begin{equation}
k_-^a=-2 \nabla^a u^+=\Omega^{-2}(\partial_-+\beta^A \partial_A)^a, \label{k-}
\end{equation}
where $\partial_\pm:=\partial_{u^\pm}$. The orientations of $u^\pm$ are chosen such that $k_+^a$ and $k_-^a$ are future-directed outgoing and incoming null vectors, respectively. From these definitions, it follows that
\begin{equation}
g_{ab}k_+^a k_-^b=-2\Omega^{-2}.
\end{equation}
These null gradient vector fields are tangent to affinely parametrized null geodesics and satisfy
\begin{equation}
k_\pm^a \nabla_a k_\pm^b=0. 
\end{equation}
For later convenience, we introduce the rescaled null vectors 
\begin{equation}
K_+^a:=\Omega k_+^a=\Omega^{-1}(\partial_+)^a, \label{K+}
\end{equation}
\begin{equation}
K_-^a:=\Omega k_-^a=\Omega^{-1}(\partial_-+\beta^A\partial_A)^a. \label{K-}
\end{equation}
They satisfy 
\begin{equation}
g_{ab}K_+^aK_-^b=-2. 
\end{equation}
They also satisfy the non-affinely parametrized null geodesic equation,
\begin{eqnarray}
K_\pm^a \nabla_a K^b_\pm=2\omega_\pm K^b_\pm,
\end{eqnarray}
where we define $\omega_\pm$ by 
\begin{eqnarray}
\omega_\pm:=\frac{1}{2}K^a_{\pm}\nabla_a \ln \Omega. 
\label{defomega}
\end{eqnarray}
The metric can also be written as 
\begin{eqnarray}
g_{ab}=-\frac{1}{2}(K_{+a}K_{-b}+K_{-a}K_{+b})+\gamma_{ab}. 
\end{eqnarray}

We also introduce the following geometrical quantities: 
\begin{eqnarray}
\theta_{\pm ab}:={\gamma_a}^c {\gamma_b}^d \nabla_c K_{\pm d}.
\label{deftheta}
\end{eqnarray}
The tensors $\theta_{\pm ab}$ are the null second fundamental forms, whose traces and traceless parts give the null expansions and shears, respectively.
We also define the vectors $v_{\pm a}$ by
\begin{eqnarray}
v_{\pm a}: =-\frac{1}{2} {\gamma_a}^b K_\mp^c \nabla_c K_{\pm b}=-{\cal D}_a\ln \Omega \mp \zeta_a,
\label{defv}
\end{eqnarray}
where ${\cal D}_a$ is the covariant derivative with respect to $\gamma_{ab}$ and 
\begin{eqnarray}
\zeta_a:= \frac{1}{2}{\gamma_a}^b K_-^c \nabla_b K_{+c}=-\frac{1}{2}{\gamma_a}^b K_+^c \nabla_b K_{-c}. 
\label{defzeta}
\end{eqnarray}
With the above definitions, $\nabla_a K_{\pm b}$ is decomposed as
\begin{eqnarray}
\nabla_a K_{\pm b}=\theta_{\pm ab}+K_{\pm a }v_{\pm b}-\omega_\pm K_{\mp a}K_{\pm b}
+\omega_\mp K_{\pm a} K_{\pm b} \mp \zeta_a K_{\pm b}. 
\end{eqnarray}

We denote the Lie derivative along $K_\pm^a$ by $\mbox \pounds_{K_\pm}$.
We define $R_{++}:=R_{ab}K_+^aK_+^b$.
The evolution equation for the outgoing null expansion $\theta_+ (:= \theta_{+a}^{\,a})$ along the outgoing null direction, namely the Raychaudhuri equation,
\begin{eqnarray}
\mbox \pounds_{K_+} \theta_+-2\omega_+ \theta_+ +\frac{1}{2}\theta_+^2
+2\rho_{+{\rm gw}}=-R_{++} \label{Ray}
\end{eqnarray}
plays a central role in the following analysis, where 
\begin{eqnarray}
\rho_{+{\rm gw}}:=\frac{1}{2}\tilde \theta_{+ab} \tilde \theta_{+}^{~ab} 
\end{eqnarray}
is interpreted as the incoming energy flux of gravitational waves, and $\tilde \theta_{\pm ab}:=\theta_{\pm ab}-(1/2)\gamma_{ab}\theta_\pm$ is the shear. 
Using Eqs.  \eqref{K+} and \eqref{defomega}, we can rewrite Eq.~\eqref{Ray} as
\begin{eqnarray}
\partial_{+}(\Omega^{-1}\theta_+)=-\frac{1}{2}\theta_+^2
-2\rho_{+{\rm gw}}-R_{++}.
\label{Ray-coordinate}
\end{eqnarray}
The evolution equation for the incoming null expansion $\theta_- (:= \theta_{-a}^{\,a})$ along the incoming null direction is obtained similarly: 
\begin{eqnarray}
\mbox \pounds_{K_-} \theta_- -2\omega_- \theta_- +\frac{1}{2}\theta_-^2
+\tilde \theta_{-ab} \tilde \theta_-^{~ab}=-R_{--}, \label{Ray-}
\end{eqnarray}
where we defined $R_{--}:=R_{ab}K^a_-K_-^b$. 
To study the evolution of $\theta_{+ab}$ in the incoming null direction and of $\theta_{-ab}$ in the outgoing null direction, we first use the following decomposition of the Riemann tensor:
\begin{eqnarray}    
{\gamma_a}^c {\gamma_b}^dR_{cedf}\gamma^{ef}&=&
{\cal R}_{ab}
-\frac{1}{2}(\theta_{+ ac} {\theta_{- b}}^{c}+\theta_{- ac} {\theta_{+ b}}^{c})
-{\cal D}_a v_{\pm b}-{\cal D}_b v_{ \pm a}
\nonumber \\
&& \quad
-2\omega_\mp \theta_{\pm ab}+2 v_{\pm a} v_{\pm b}
-{\gamma_a}^c {\gamma_b}^d \nabla_{K_\mp}\theta_{\pm cd},
\label{ggRg}
\end{eqnarray}
where ${\cal R}_{ab}:={\gamma_a}^c {\gamma_b}^d R_{cd}$.
Using the Gauss equation
\begin{eqnarray}
{}^{(2)}R_{abcd}={\gamma_a}^e {\gamma_b}^f {\gamma_c}^g {\gamma_d}^h R_{efgh}+\frac{1}{2}
(-\theta_{+ac}\theta_{-bd}-\theta_{-ac}\theta_{+bd}+\theta_{+bc}\theta_{-ad}+\theta_{-bc}\theta_{+ad}),
\end{eqnarray}
where ${}^{(2)}R_{abcd}$ is the Riemann tensor associated with $\gamma_{ab}$, 
the left-hand side of Eq.~\eqref{ggRg} becomes
\begin{equation}
{\gamma_a}^c {\gamma_b}^dR_{cedf}\gamma^{ef}={}^{(2)}R_{ab}-\frac{1}{2}(-\theta_{+ab} \theta_--\theta_{-ab}\theta_+
+{\theta_{+a}}^{c}\theta_{-bc}+{\theta_{-a}}^{c}\theta_{+bc} ).
\label{ggRg2}
\end{equation}
Combining Eqs.~\eqref{ggRg} and \eqref{ggRg2} and rewriting the result in terms of the Lie derivative along $K_\mp^a$, we obtain 
\begin{eqnarray}
\mbox \pounds_{K_\mp} \theta_{\pm ab}
 & = &\theta_{- ac} {\theta_{+ b}}^{c}+\theta_{- bc} {\theta_{+ a}}^{c}-{}^{(2)}R_{ab}
+{\cal R}_{ab}
-{\cal D}_a v_{ \pm b}-{\cal D}_b v_{ \pm a}-2\omega_\mp \theta_{\pm ab}
\nonumber \\
&&\quad
+2v_{\pm a} v_{\pm b}
 -\frac{1}{2}(\theta_{+ ab}\theta_-+\theta_{- ab} \theta_+)
\pm 2\left(K_{\mp a}\theta_{\pm bc} \zeta^c + K_{\mp b}\theta_{\pm ac} \zeta^c
\right).
\label{ggLt}
\end{eqnarray}
Since ${}^{(2)}R_{ab}=(1/2){}^{(2)}R \gamma_{ab}$ holds in two dimensions, the trace part of Eq.~\eqref{ggLt} becomes
\begin{eqnarray}
\Omega^{-1} \mbox \pounds_{K_\mp}(\Omega \theta_\pm)=-{}^{(2)}R+{\cal R}-2{\cal D}_a v_\pm^{~a}
+2v_{\pm a}v_\pm^{~a}-\theta_+ \theta_-, \label{crossfocus}
\end{eqnarray}
where ${\cal R}:=\gamma^{ab}{\cal R}_{ab}$.
The traceless part gives the variation of the gravitational-wave energy flux associated with the outgoing shear in the incoming null direction: 
\begin{eqnarray}
\Omega^{-2} \mbox \pounds_{K_-} (\Omega^2 \rho_{+{\rm gw}})+\theta_- \rho_{+{\rm gw}}
=S_{+ab}\tilde \theta_+^{~ab},\label{gwevol}
\end{eqnarray}
where 
\begin{eqnarray}
S_{+ab}:={\cal R}_{ab}-{\cal D}_a v_{+b}-{\cal D}_b v_{+a}+2v_{+a} v_{+b}-\frac{1}{2}\theta_+ \theta_{-ab}. 
\label{S+ab}
\end{eqnarray}
Note that only the traceless part of $S_{+ab}$ contributes to Eq.~\eqref{gwevol}.

The contracted Bianchi identity gives the following evolution equation for $R_{++}$ in the incoming null direction: 
\begin{eqnarray}
\Omega^{-2}\mbox \pounds_{K_-}(\Omega^2R_{++})=-\theta_-R_{++}+S_+, \label{bianchi}
\end{eqnarray}
where 
\begin{eqnarray}
S_+:=-\mbox \pounds_{K_+}{\cal R}-\theta_+(R_{+-}+{\cal R})
+2{\cal D}^a {\cal R}_{+a}-2 \tilde \theta^{~ab}_+{\cal R}_{ab}+2{\cal R}_{+a}(\zeta^a-v_-^{~a}-2v_+^{~a}),
\label{S+}
\end{eqnarray}
and $R_{+-}:=R_{ab}K^a_+K^b_-$ and ${\cal R}_{\pm a}:={\gamma_a}^bR_{bc}K^c_\pm$.

Using Eq. \eqref{K-},
we can rewrite Eqs.~\eqref{gwevol} and \eqref{bianchi} in coordinate form as
\begin{align}
\partial_- (\Omega^2 \rho_{+{\rm gw}})&=-\Omega^3  \theta_- \rho_{+{\rm gw}} -\beta^A \partial_A(\Omega^2\rho_{+{\rm gw}}) +\Omega^3 S_{+ab} \tilde \theta_+^{~ab},
\label{terhogw} \\
\partial_- (\Omega^2 R_{++})&=-\Omega^3  \theta_- R_{++} -\beta^A \partial_A(\Omega^2R_{++})  +\Omega^3 S_+.
\label{teR++}
\end{align}
The first two terms on the right-hand sides of
Eqs.~\eqref{terhogw} and \eqref{teR++} describe conservative transport
of the weighted fluxes $\Omega^2\rho_{+{\rm gw}}$ and $\Omega^2R_{++}$
along $K_-^a$, while the third terms represent source contributions.
The first terms account for the convergence or divergence
of the incoming null congruence.
The second terms describe angular advection associated with
the shift vector $\beta^A$, arising from the difference between
the directions of $K_-^a$ and $\partial_-$, as in Eq.~\eqref{K-}.

The source terms $S_{+ab}\tilde\theta_+^{~ab}$ and $S_+$
describe departures from this conservative transport.
Although $S_+$ is a linear combination of Ricci tensor components and their derivatives, it does not include $R_{++}$.
Hence, $S_+$ represents contributions from other components
of the energy-momentum tensor and their derivatives through
the Einstein equation, as reflected in Eq.~\eqref{S+}.
The gravitational-wave source $S_{+ab}\tilde\theta_+^{~ab}$
includes contributions from the geometry of the null foliation
and from matter through ${\cal R}_{ab}$,
as shown in Eq.~\eqref{S+ab}.
These couplings can enhance or reduce the incoming
fluxes relative to conservative transport, depending
on the signs of the source terms.

We also need evolution equations for ${}^{(2)}R$ and $\zeta_a$ to examine the consistency of the order estimates in Sec.~\ref{sec:order-consistency}.
A direct calculation gives the evolution equation for ${}^{(2)}R$ in the outgoing null direction: 
\begin{eqnarray}
\mbox \pounds_{K_+}{}^{(2)}R=-\theta_+ {}^{(2)}R+2\Omega^{-1}[{\cal D}^a{\cal D}^b(\Omega \theta_{+ab})-{\cal D}^2 (\Omega\theta_+) ].   \label{2Ricci}
\end{eqnarray}
Using the Codazzi identity, ${\cal R}_{\pm a}={\gamma_a}^b(\nabla^d\nabla_b-\nabla_b\nabla^d)K_{\pm d}$, we obtain the evolution equations for $\zeta_a$:
\begin{eqnarray}
\mbox \pounds_{K_\pm}\zeta_a &=& \mp{\cal R}_{\pm a}-\theta_\pm\zeta_a
\nonumber \\
&&\quad\pm\Bigl[{\cal D}^b\theta_{\pm ab}-{\cal D}_a(\theta_\pm+2\omega_\pm)-2\omega_\pm{\cal D}_a\ln\Omega +\theta_{\pm ab}{\cal D}^b\ln\Omega-2K_{\pm a}\zeta_b\zeta^b\Bigr]. \label{null-AB}
\end{eqnarray}

For the subsequent analysis, we derive coordinate expressions for $\zeta_a$ and $\theta_{\pm ab}$. We define their components in the coordinates $x^A$ on the two-surfaces by
\begin{eqnarray}
\zeta_a &=& \zeta_A\bigl[(dx^A)_a-\beta^A(du^-)_a\bigr],
\\
\theta_{\pm ab} &=& \theta_{\pm AB}\bigl[(dx^A)_a-\beta^A(du^-)_a\bigr]\bigl[(dx^B)_b-\beta^B(du^-)_b\bigr].
\label{coordinate-components}
\end{eqnarray}
Using the definitions of $\zeta_a$ and $\theta_{\pm ab}$ in Eqs.~\eqref{defzeta} and \eqref{deftheta}, respectively, we obtain
\begin{eqnarray}
\zeta_A=-\frac{1}{4}\Omega^{-2} \gamma_{AB}\partial_+ \beta^B,\label{zeta-beta}
\end{eqnarray}
\begin{eqnarray}
\theta_{+AB}=\frac{1}{2\Omega}\partial_+ \gamma_{AB},
\label{+ABtheta}
\end{eqnarray}
and 
\begin{eqnarray}
\theta_{-AB}=\frac{1}{2\Omega}\left(\partial_-\gamma_{AB}+\beta^C\partial_C\gamma_{AB}+\gamma_{CB}\partial_A\beta^C+\gamma_{AC}\partial_B\beta^C\right).
\end{eqnarray}
The trace of Eq.~\eqref{+ABtheta} becomes 
\begin{eqnarray}
\theta_+=\frac{1}{2\Omega}\gamma^{AB}\partial_+ \gamma_{AB}
=\frac{1}{\Omega}\partial_+ \ln {\sqrt {\det (\gamma_{AB})}}
. \label{theta-phi}
\end{eqnarray}

\section{Trapped surface formation} \label{TSF}

Using the double-null framework introduced in Sec.~\ref{basics}, we study trapped surface formation caused by a short incoming pulse of gravitational radiation and matter.

\subsection{Setup and initial data}

We take the spacetime to be Minkowski within a causal diamond ${\cal M}_0$, bounded by the future light cone of an event and the past light cone of a later, timelike-separated event.
In ${\cal M}_0$, we choose spherical double-null coordinates\footnote{In ${\cal M}_0$, the usual Minkowski time and radial coordinates are $t=u^++u^-$ and $r=u^+-u^-\geq0$, respectively. Thus, $-dt^2+dr^2=-4du^+du^-$.} in which the metric takes the form
\begin{eqnarray}
g_0=-4du^+du^-+\gamma_{0AB}dx^Adx^B,
\end{eqnarray}
where 
\begin{eqnarray}
\gamma_{0AB}=(u^+-u^-)^2 \sigma_{AB}
\label{Minksph}
\end{eqnarray}
and $\sigma_{AB} dx^Adx^B$ is the metric on the unit round sphere. 
We introduce a length scale $r_0>0$ to specify the size of ${\cal M}_0$.
We choose the coordinates $(u^+,u^-)$ so that the future and past vertices of ${\cal M}_0$ are located at $(u^+,u^-)=(0,0)$ and $(u^+,u^-)=(-r_0,-r_0)$, respectively.
Thus, including its boundary, ${\cal M}_0$ is specified by $-r_0\leq u^-\leq u^+\leq0$.
The spatial edge of ${\cal M}_0$, located at $(u^+,u^-)=(0,-r_0)$, forms a two-sphere of radius $r_0$, denoted by $S_0$.

In ${\cal M}_0$, the outgoing and incoming null expansions, denoted by $\theta_{0\pm}$, are computed as 
\begin{eqnarray}
\theta_{0 \pm}=\pm 2\frac{1}{u^+-u^-}.
\label{Minkexpansions}
\end{eqnarray}
At $S_0$, they are evaluated as
\begin{eqnarray}
\theta_{0\pm}|_{S_0}=\pm\frac{2}{r_0}.
\end{eqnarray}

Let $\delta>0$ denote the width of the incoming pulse in the $u^+$ coordinate.
The initial data for the gravitational waves and matter are given on the initial null hypersurface ${\cal N}_0$, which extends outward from $S_0$ and is specified by $u^-=-r_0$ and $0 \leq u^+ \leq \delta$.
With this setup, no gravitational waves or matter propagate in the Minkowski causal diamond ${\cal M}_0$.

We choose a parameter $0<r_f<r_0$ and seek conditions on the initial data that ensure that the two-surface $T$ at $(u^+,u^-)=(\delta,-r_f)$ is trapped.
To this end, we follow the inward propagation of the gravitational waves and matter through the region ${\cal M}_1$, specified by $-r_0\leq u^-\leq-r_f$ and $0\leq u^+\leq\delta$, which lies to the future of and outside ${\cal M}_0$.
We denote the null boundary of ${\cal M}_1$ at $u^-=-r_f$ by ${\cal N}_f$.
We keep $r_f$ fixed as $\delta$ is made small. In the following, all relations and conditions are understood pointwise in the angular coordinates $x^A$.

On ${\cal N}_0$, we adopt the same $\delta$-scaling for the initial 
gravitational-wave flux as in Christodoulou's short-pulse 
construction~\cite{Christodoulou2009}, so that the initial energy flux of 
gravitational waves $\rho_{+{\rm gw}}^{{\cal N}_0}:=\rho_{+{\rm gw}}(u^+,-r_0)$ 
satisfies
\begin{eqnarray}
\rho_{+{\rm gw}}^{{\cal N}_0}=O(\delta^{-1}).
\label{initialgw}
\end{eqnarray}
We impose the same scaling on the initial incoming energy flux of matter
$R_{++}^{{\cal N}_0}:=R_{++}(u^+,-r_0)$, so that gravitational waves and matter contribute at the same order:
\begin{eqnarray}
R_{++}^{{\cal N}_0}=O(\delta^{-1}).
\label{initialRicci}
\end{eqnarray}
The bounds in Eqs.~\eqref{initialgw} and
\eqref{initialRicci} hold uniformly on ${\cal N}_0$,
with constants independent of $\delta$, $u^+$,
and the angular coordinates.

\subsection{Flux transport and formation criterion}

We first note that Eq.~\eqref{Ray-coordinate} governs the evolution of the outgoing expansion $\theta_+$ as 
\begin{eqnarray}
\partial_{+}(\Omega^{-1}\theta_+)=-\frac{1}{2}\theta_+^2 -2\rho_{+{\rm gw}}-R_{++} \leq -2\rho_{+{\rm gw}}-R_{++}. \label{Ray2}
\end{eqnarray}
Integrating this equation with respect to $u^+$ from $0$ to $\delta$ on the final null hypersurface ${\cal N}_f$ gives 
\begin{eqnarray}
(\Omega^{-1}\theta_+)(\delta,-r_f) \leq (\Omega^{-1}\theta_+)(0,-r_f)-\int_0^\delta \Bigl(2\rho_{+{\rm gw}}^{{\cal N}_f}+R_{++}^{{\cal N}_f} \Bigr)du^+,
\end{eqnarray}
where $\rho_{+{\rm gw}}^{{\cal N}_f}:=\rho_{+{\rm gw}}(u^+,-r_f)$ and $R_{++}^{{\cal N}_f}=R_{++}(u^+,-r_f)$.  
At $(u^+,u^-)=(0,-r_f)$ on the boundary of the
Minkowski region ${\cal M}_0$, $\Omega=1$ and
Eq.~\eqref{Minkexpansions} gives
$(\Omega^{-1}\theta_+)(0,-r_f)=2/r_f$.
Thus, if the condition
\begin{eqnarray}
\int_0^\delta \Bigl(2\rho_{+{\rm gw}}^{{\cal N}_f}+R_{++}^{{\cal N}_f} \Bigr)du^+ \geq \frac{2}{r_f} \label{gcon0}
\end{eqnarray}
holds, $\theta_+ (\delta,-r_f)$ becomes negative.\footnote{Equality in Eq.~\eqref{gcon0} also suffices, because 
the term omitted in Eq.~\eqref{Ray2}, $\frac12\int_0^\delta\theta_+^2du^+$, is strictly positive.} 
This is a sufficient condition on the final gravitational-wave and matter fluxes for $\theta_+<0$ on $T$. 

Our goal is to derive a sufficient condition on the initial gravitational-wave flux $\rho_{+{\rm gw}}^{{\cal N}_0}:=\rho_{+{\rm gw}}(u^+,-r_0)$ 
and the initial matter flux $R_{++}^{{\cal N}_0}:=R_{++}(u^+,-r_0)$ that ensures condition~\eqref{gcon0} holds. 
To this end, we examine the evolution of the gravitational-wave flux $\rho_{+{\rm gw}}$ and the matter flux $R_{++}$ in the incoming null direction. 
For convenience, we rewrite the flux evolution equations
derived in Sec.~\ref{basics}:
\begin{align*}
\partial_- (\Omega^2 \rho_{+{\rm gw}})&=-\Omega^3  \theta_- \rho_{+{\rm gw}} -\beta^A \partial_A(\Omega^2\rho_{+{\rm gw}}) +\Omega^3 S_{+ab} \tilde \theta_+^{~ab},
\tag{\ref{terhogw}} \\
\partial_- (\Omega^2 R_{++})&=-\Omega^3  \theta_- R_{++} -\beta^A \partial_A(\Omega^2R_{++})  +\Omega^3 S_+.
\tag{\ref{teR++}}
\end{align*}

The first terms on the right-hand sides of
Eqs.~\eqref{terhogw} and \eqref{teR++} describe the concentration
of energy flux as the incoming null congruence converges.
Assuming that the second and third terms, namely the angular advection and source terms, are subleading, we use this 
amplification to derive a sufficient condition on the initial fluxes for trapped surface formation. 
To this end, we make the following additional assumptions. 

First, the lapse function and the incoming expansion are assumed to remain 
close to their Minkowski values,
\begin{eqnarray}
\Omega= 1+ O(\delta^{1/2})
\label{lapse-assumption}
\end{eqnarray} 
and 
\begin{eqnarray}
\theta_-= \theta_{0-}+ O(\delta^{1/2}). 
\label{theta-minus-assumption}
\end{eqnarray} 
Here, we extend $\theta_{0-}$ to ${\cal M}_1$
as a reference function using the coordinate expression
in Eq.~\eqref{Minkexpansions}.

Second, the angular advection and source terms are assumed to be subleading:
\begin{align}
\beta^A\partial_A(\Omega^2\rho_{+{\rm gw}})
&=O(\delta^{-1/2}),
\label{flux-remainder-bounds}\\
\beta^A\partial_A(\Omega^2R_{++})
&=O(\delta^{-1/2}),
\label{flux-remainder-bounds-matter}\\
\Omega^3S_{+ab}\tilde\theta_+^{~ab}
&=O(\delta^{-1/2}),
\label{flux-source-bounds}\\
\Omega^3S_+
&=O(\delta^{-1/2}).
\label{flux-source-bounds-matter}
\end{align}
These bounds, together with the estimates for $\Omega$ and
$\theta_-$ above, are assumed to hold uniformly throughout
${\cal M}_1$, with constants independent of $\delta$.
These assumptions are examined
for consistency with the evolution equations
at the level of order estimates
in the next section.

Under these assumptions, Eqs.~\eqref{terhogw} and \eqref{teR++} become
\begin{eqnarray}
\partial_-(\Omega^2\rho_{+{\rm gw}})
&=&
-\Omega\theta_-(\Omega^2\rho_{+{\rm gw}})
+O(\delta^{-1/2}),
\label{evleqs}\\
\partial_-(\Omega^2R_{++})
&=&
-\Omega\theta_-(\Omega^2R_{++})
+O(\delta^{-1/2}).
\label{evleqs-matter}
\end{eqnarray}
Equations~\eqref{evleqs} and \eqref{evleqs-matter} imply that the bounds 
$\Omega^2\rho_{+{\rm gw}},\,\Omega^2R_{++}=O(\delta^{-1})$, which hold on 
${\cal N}_0$ by the initial flux estimates, are maintained throughout 
${\cal M}_1$.\footnote{
At fixed $u^+$ and $x^A$, let $Q$ denote either
$\Omega^2\rho_{+{\rm gw}}$ or $\Omega^2R_{++}$,
and let $E=O(\delta^{-1/2})$ denote the corresponding
remainder in Eqs.~\eqref{evleqs} and \eqref{evleqs-matter}.
Their integration gives
\begin{align*}
Q(u^-)
&=\exp\left[-\int_{-r_0}^{u^-}
(\Omega\theta_-)(s)\,ds\right]Q(-r_0)+\int_{-r_0}^{u^-}
\exp\left[-\int_t^{u^-}(\Omega\theta_-)(s)\,ds\right]
E(t)\,dt.
\end{align*}
Since $\Omega\theta_-$ is uniformly bounded and the
integration interval has a fixed length, both exponential
factors are bounded independently of $\delta$.
The first and second terms are therefore
$O(\delta^{-1})$ and $O(\delta^{-1/2})$, respectively.}
Since $\Omega=1+O(\delta^{1/2})$, this also gives
\begin{equation}
\rho_{+{\rm gw}}=O(\delta^{-1}),\qquad
R_{++}=O(\delta^{-1})
\label{propagated-flux-bounds}
\end{equation}
throughout ${\cal M}_1$. Using this bound and $\Omega\theta_-=\theta_{0-}+O(\delta^{1/2})$,
we absorb the coefficient corrections into the $O(\delta^{-1/2})$ remainders and rewrite 
Eqs.~\eqref{evleqs} and \eqref{evleqs-matter} as
\begin{eqnarray}
\partial_-(\Omega^2\rho_{+{\rm gw}})
&=&
-\theta_{0-}(\Omega^2\rho_{+{\rm gw}})
+O(\delta^{-1/2}),
\label{evleqs-leading}\\
\partial_-(\Omega^2R_{++})
&=&
-\theta_{0-}(\Omega^2 R_{++})
+O(\delta^{-1/2}).
\label{evleqs-leading-matter}
\end{eqnarray}
Integrating Eqs.~\eqref{evleqs-leading} and \eqref{evleqs-leading-matter} over
$-r_0\leq u^-\leq-r_f$ with Eq.~\eqref{Minkexpansions}, and using
$0\leq u^+\leq\delta$ and $\Omega=1+O(\delta^{1/2})$,
we obtain
\begin{eqnarray}
\rho_{+{\rm gw}}^{{\cal N}_f}= \Bigl(\frac{r_0}{r_f} \Bigr)^2 \rho_{+{\rm gw}}^{{\cal N}_0}  + O(\delta^{-1/2}),
\qquad
R_{++}^{{\cal N}_f}=\Bigl(\frac{r_0}{r_f} \Bigr)^2R_{++}^{{\cal N}_0} + O(\delta^{-1/2}).
\end{eqnarray}

The condition~\eqref{gcon0} for trapped surface formation on ${\cal N}_f$ can then be expressed 
in terms of the fluxes on ${\cal N}_0$, $\rho_{+{\rm gw}}^{{\cal N}_0}$ and $R_{++}^{{\cal N}_0}$, as 
\begin{eqnarray}
\int_0^\delta \Bigl(2\rho_{+{\rm gw}}^{{\cal N}_0}+R_{++}^{{\cal N}_0} \Bigr)du^+ > \frac{2r_f}{r_0^2}  +O(\delta^{1/2}). \label{initialcriterion}
\end{eqnarray}
The inequality is understood to hold at every angle with a margin larger than a uniform bound on
the indicated $O(\delta^{1/2})$ correction. Therefore, under condition~\eqref{initialcriterion},
$\theta_+<0$ on $T$. For the ingoing null expansion of $T$, the assumed estimate in
Eq.~\eqref{theta-minus-assumption} gives
$\theta_-(\delta,-r_f)=-2/(\delta+r_f)+O(\delta^{1/2})<0$
for sufficiently small $\delta$.

We summarize the setup, assumptions, and conclusion of the preceding 
argument. 
We first recall the setup: a Minkowski region ${\cal M}_0$, a fixed 
$0<r_f<r_0$ independent of $\delta$, and initial data on ${\cal N}_0$ 
satisfying Eqs.~\eqref{initialgw} and \eqref{initialRicci}.
We assume that a regular spacetime with a regular double-null foliation exists 
throughout ${\cal M}_1$, and that the bounds in 
Eqs.~\eqref{lapse-assumption}--\eqref{flux-source-bounds-matter} hold 
uniformly there, with constants independent of $\delta$. 
Under these assumptions, if the initial flux condition~\eqref{initialcriterion} 
holds at every angle with a margin exceeding the indicated $O(\delta^{1/2})$ 
correction uniformly in angle, then the two-surface $T$ at 
$(u^+,u^-)=(\delta,-r_f)$ is trapped for sufficiently small $\delta$.

Condition~\eqref{initialcriterion} can be rearranged as 
\begin{eqnarray}
r_f < r_0^2 \int_0^\delta \rho^{{\cal N}_0} \, du^+   +O(\delta^{1/2}),\label{radiuscriterion}
\end{eqnarray}
where $\rho^{{\cal N}_0}$ is defined by 
\begin{eqnarray}
\rho^{{\cal N}_0} := \rho_{+{\rm gw}}^{{\cal N}_0}+ \frac12 R_{++}^{{\cal N}_0}
\end{eqnarray}
and is interpreted as the total incoming energy flux.
Note that $\rho^{{\cal N}_0}$ includes not only the energy flux of gravitational waves
but also that of matter.
This additive structure also appears in sufficient conditions
on the initial flux for trapped surface formation
in specific Einstein--matter systems~\cite{Yu:2011qy,AnAthanasiou2020,ZhaoHilditchValienteKroon2023,AthanasiouMondalYau2023,Shen:2025ztd}.
Here, we express the matter contribution through the Ricci tensor
and derive the flux criterion without specifying a matter model,
under the estimates assumed above throughout ${\cal M}_1$.
It is physically natural for the total energy flux
to be the sum of the contributions from the individual fields.
In vacuum, condition~\eqref{radiuscriterion} agrees at leading order
with Christodoulou's trapped surface formation criterion~\cite{Christodoulou2009},
also discussed by Klainerman and Rodnianski~\cite{KlainermanRodnianski2012}
and Bini and Esposito~\cite{BiniEsposito2017}.

It is useful to examine the dimensions of the right-hand
side of Eq.~\eqref{radiuscriterion}.
The quantity $r_0^2\int_0^\delta \rho^{{\cal N}_0}\,du^+$
is the length scale associated with the incoming energy flux
measured on the initial null surface ${\cal N}_0$.
The criterion~\eqref{radiuscriterion} states that,
under the assumptions above,
a trapped surface forms if $r_f$ is smaller than this scale
at every angle, with a margin exceeding the indicated correction.
In this sense, the criterion reflects the physical idea
of energy concentration underlying the hoop conjecture~\cite{hoop}.
This connection is qualitative:
the present condition is imposed pointwise in angle,
and we do not identify it with an inequality involving
total mass and circumference.

Our aim is to describe the formation of a trapped surface
during gravitational collapse, rather than cases in which
the initial null hypersurface ${\cal N}_0$ already contains
trapped sections. We therefore choose the incoming pulse so 
that the outgoing expansion remains positive on the initial null hypersurface ${\cal N}_0$. To see that this is compatible with the formation condition~\eqref{initialcriterion}, we integrate Eq.~\eqref{Ray2} on ${\cal N}_0$. At $(u^+,u^-)=(\delta,-r_0)$, using $\theta_+=O(1)$, we obtain
\begin{eqnarray}
(\Omega^{-1}\theta_+)(\delta,-r_0)
=\frac{2}{r_0}- 2\int_0^\delta \rho^{{\cal N}_0} du^++O(\delta).
\label{initialsurface}
\end{eqnarray}
Choosing $\int_0^\delta \rho^{{\cal N}_0}du^+<1/r_0$, with a margin larger than the $O(\delta)$ correction, makes the outgoing expansion positive at $u^+=\delta$.
To ensure positivity also on all earlier sections of ${\cal N}_0$,
it suffices, for example, to assume that the \emph{combined} focusing
term satisfies
\begin{eqnarray}
\rho^{{\cal N}_0}\geq 0 
\qquad (0\leq u^+\leq\delta).
\label{combined-focus-initial}
\end{eqnarray}
Indeed, the exact Raychaudhuri equation~\eqref{Ray2} then implies that
$\Omega^{-1}\theta_+$ is non-increasing in $u^+$ on ${\cal N}_0$,
so its positivity at $u^+=\delta$ implies positivity on every
preceding section. More generally, it is enough
that the integral of $\rho^{{\cal N}_0}$
from any earlier $u^+$ to $\delta$ be non-negative, since the omitted
$\theta_+^2/2$ term is also non-negative.
Under this sufficient condition, the initial null hypersurface is
untrapped, while the same flux can satisfy the formation
condition~\eqref{initialcriterion}, whose leading threshold is
$2r_f/r_0^2<2/r_0$. At leading order, inward transport amplifies the integrated
flux by a factor $(r_0/r_f)^2$, whereas the threshold in
Eq.~\eqref{gcon0} increases only by a factor $r_0/r_f$.
This describes the dynamical formation of a trapped surface from initially untrapped data, as the incoming gravitational radiation and matter become concentrated at smaller radii during collapse.

\section{Order assumptions and their consistency}
\label{sec:order-consistency}

In the previous section, we assumed $\Omega-1=O(\delta^{1/2})$ and 
$\theta_--\theta_{0-}=O(\delta^{1/2})$, as stated in 
Eqs.~\eqref{lapse-assumption} and \eqref{theta-minus-assumption}, together 
with the bounds on the angular advection and source terms in 
Eqs.~\eqref{flux-remainder-bounds}--\eqref{flux-source-bounds-matter}, which 
ensure approximately conservative transport of the incoming gravitational-wave 
and matter fluxes. 
All these estimates were assumed to hold uniformly throughout ${\cal M}_1$, 
with constants independent of $\delta$. 
In this section, we identify a more detailed hierarchy of orders for the 
geometric quantities, the Ricci tensor, and their derivatives throughout 
${\cal M}_1$. 
This hierarchy includes the flux bounds in 
Eq.~\eqref{propagated-flux-bounds}, obtained in the previous section from the 
initial flux estimates and Eqs.~\eqref{evleqs} and \eqref{evleqs-matter}. 
Under this hierarchy, we obtain the bounds in 
Eqs.~\eqref{flux-remainder-bounds}--\eqref{flux-source-bounds-matter} and 
examine the consistency of the assumed orders with the evolution equations.

\subsection{Hierarchy of orders}

We first state the order assumptions for the relevant variables. 
The assumptions are given in Sec.~\ref{ssSec:OA}, and in 
Sec.~\ref{ssSec:ConsOA} we derive the further orders that follow from them and 
verify the bounds in 
Eqs.~\eqref{flux-remainder-bounds}--\eqref{flux-source-bounds-matter} used in 
the formation argument of the previous section.

\subsubsection{Order assumptions}
\label{ssSec:OA}

We begin with the following bounds in powers of $\delta$:
\begin{eqnarray}
&&\omega_+ = O(\delta^{-1/2}), \quad 
\omega_-=O(\delta^{1/2}), \quad 
\zeta_a = O(\delta^0), \quad 
v_{\pm a} =O(\delta^0), \quad
\theta_\pm=  O(\delta^0), 
\nonumber \\ 
&& 
\tilde \theta_{+ab}= O(\delta^{-1/2}),  \quad
\tilde \theta_{-ab}= O(\delta^{1/2}), \quad 
\Omega -1 = O(\delta^{1/2}), \quad 
\Delta \theta_-:= \theta_- - \theta_{0-}= O(\delta^{1/2}). 
\label{BS}
\end{eqnarray}
These conditions allow the incoming gravitational waves
to have a large shear $\tilde \theta_{+ab}$,
while $\Omega$ and $\theta_-$ remain close to their Minkowski values.\footnote{
Since $\omega_+$ is of the same order as
$\tilde\theta_{+ab}$, one might expect it to be equally relevant.
However, the underlying geometrical quantity is $\Omega$,
whose deviation from the Minkowski value is of order
$\delta^{1/2}$, while $\omega_+$ is proportional to its
logarithmic derivative along the $u^+$ direction.
Equation~\eqref{defomega} gives
$\partial_+\Omega=2\Omega^2\omega_+$.
For bounded $\Omega$, the assumed bound
$\omega_+=O(\delta^{-1/2})$ therefore gives a change
of at most $O(\delta^{1/2})$ across the pulse width $\delta$.
Furthermore, when the Raychaudhuri equation is written
in terms of $\Omega^{-1}\theta_+$, as in
Eq.~\eqref{Ray-coordinate}, the explicit $\omega_+$ term
is absorbed into the derivative on the left-hand side
and does not appear as a separate focusing source.
}
In the vacuum limit, these orders coincide with the pointwise ($L^\infty$) 
bootstrap bounds of Ref.~\cite{KlainermanRodnianski2012}.
The shear bound in Eq.~\eqref{BS} implies
$\rho_{+{\rm gw}}=O(\delta^{-1})$ throughout ${\cal M}_1$,
and thus includes the initial flux bound in
Eq.~\eqref{initialgw} on ${\cal N}_0$.
For tensors tangent to the $u^+,~u^-=$const. two-surfaces, the order estimates
are understood componentwise in a local orthonormal frame
with respect to the induced metric $\gamma$.

We also allow an incoming matter pulse to contribute to
trapped surface formation at the same leading order as
gravitational radiation.
We impose the following order bounds on the Ricci tensor:
\begin{eqnarray}
&&R_{++}= O(\delta^{-1}), \qquad R_{+-}= O(\delta^{-1/2}), \qquad R_{--}= O(\delta^{1/2}), \nonumber \\ 
&&{\cal R}_{+a}= O(\delta^{0}),  
\qquad {\cal R}_{ab}= O(\delta^{1/2}).
\label{RS}
\end{eqnarray}
The bound on $R_{++}$ in Eq.~\eqref{RS} includes
the initial matter flux bound in Eq.~\eqref{initialRicci}
on ${\cal N}_0$.
No independent restriction on ${\cal R}_{-a}$ is needed
for the main argument of this paper.
The hierarchy in Eq.~\eqref{RS} allows the incoming
energy flux $R_{++}$ to be dominant,
while the outgoing and transverse components are suppressed.
Thus, it characterizes matter that propagates predominantly along the incoming null direction, 
with little conversion of its flux into other components during the passage of the pulse.
Pressureless matter propagating along the null direction is expected to satisfy these conditions, 
although we do not attempt to verify this for any specific matter model here.
Verifying them for a given matter model requires solving its field equations.

We also impose the following derivative bound\footnote{
This is the order assigned by the counting of derivatives along $K_\pm$ in 
Sec.~\ref{ssSec:CD}. The bounds treated there, however, are read off from the 
evolution equations and their consistency under integration is checked, 
whereas ${\cal R}$ obeys no evolution equation here, so that 
Eq.~\eqref{ricci-trace-derivative-bound} is imposed as an order assumption.
}:
\begin{align}
\mbox \pounds_{K_+}{\cal R}&=O(\delta^{-1/2}).
\label{ricci-trace-derivative-bound}
\end{align}
For a quantity $\psi=O(\delta^s)$,
we assign the angular derivative bound
\begin{align}
{\cal D}_a\psi = O(\delta^{s-1/2}).
\label{angular-derivative-counting}
\end{align}
In particular, Eq.~\eqref{angular-derivative-counting} gives
\begin{align}
{\cal D}_a\tilde\theta_{+bc}=O(\delta^{-1}),\qquad
{\cal D}_aR_{++}=O(\delta^{-3/2}),\qquad
{\cal D}^a{\cal R}_{+a}=O(\delta^{-1/2}).
\label{flux-derivative-bounds}
\end{align}
These are the angular derivative bounds used to derive 
Eqs.~\eqref{flux-remainder-bounds}--\eqref{flux-source-bounds-matter}.
The angular derivative counting in
Eq.~\eqref{angular-derivative-counting}
corresponds to the signature assignment of
Ref.~\cite{KlainermanRodnianski2012}, where an angular
derivative raises the weight by $1/2$.\footnote{
In a second-order system, ${\cal D}^2$ may appear together with 
$\nabla_{K_+}\nabla_{K_-}$. 
The angular derivative counting adopted here, combined with the null 
derivative counting introduced in Sec.~\ref{ssSec:CD} below, assigns two 
angular derivatives and one derivative along each null direction the same 
total weight. 
This balance motivates the angular derivative assumption; it does not 
establish the corresponding estimates from initial data.
}

The bounds introduced so far suffice for the consistency of the evolution 
equations examined in Sec.~\ref{sec:evolution-consistency}. 
For the formation argument, however, the gravitational-wave source in 
Eq.~\eqref{flux-source-bounds} requires in addition
\begin{eqnarray}
\left({\cal D}_a v_{+b}\right)^{(STL)}=O(\delta^0),
\label{angularBS}
\end{eqnarray}
where $(STL)$ denotes the symmetric traceless part with respect to 
$\gamma_{ab}$,
\begin{eqnarray}
\left({\cal D}_a v_{+b}\right)^{(STL)}:=\frac12 
\left({\gamma_a}^c{\gamma_b}^d+{\gamma_a}^d{\gamma_b}^c
-\gamma_{ab}\gamma^{cd}\right){\cal D}_c v_{+d}.
\end{eqnarray}
To close the hierarchy for Eq.~\eqref{angularBS} under its own evolution 
equation, we also impose
\begin{align}
{\cal D}^2\theta_{+ab}=O(\delta^{-1}),
\label{angular-laplacian-bound}
\end{align}
which is equivalent to ${\cal D}^2\tilde\theta_{+ab}=O(\delta^{-1})$. 
The bounds in Eqs.~\eqref{angularBS} and \eqref{angular-laplacian-bound} are 
exceptions to the counting in Eq.~\eqref{angular-derivative-counting}, being 
stronger by $\delta^{1/2}$ than the orders it assigns. 
Each further angular derivative of these quantities, however, is assumed to 
lower the power of $\delta$ by $1/2$, as in 
Eq.~\eqref{angular-derivative-counting}.

\subsubsection{Consequences of the order assumptions}
\label{ssSec:ConsOA}

Here we derive the order estimates that follow from the order assumptions 
given in Sec.~\ref{ssSec:OA}. 
Integrating Eq.~\eqref{+ABtheta} over the pulse width $\delta$, with the Minkowski data at $u^+=0$ and the bound in Eq.~\eqref{BS}, 
gives
\begin{eqnarray}
\gamma_{ab} - \gamma_{0ab} = O(\delta^{1/2}),
\label{angular-metric-estimate}
\end{eqnarray}
where $\gamma_{0ab}$ is the angular part of the Minkowski metric, extended to 
${\cal M}_1$ as a reference two-metric using the coordinate expression in 
Eq.~\eqref{Minksph}.\footnote{
In deriving Eq.~\eqref{angular-metric-estimate},
we use the bounds in the local orthonormal frame
to control the relative rate of change of the squared lengths
of vectors with fixed angular coordinate components.
}

Equation~\eqref{zeta-beta}, together with the bounds in Eq.~\eqref{BS} and the 
metric estimate in Eq.~\eqref{angular-metric-estimate}, gives 
$\partial_+\beta^A=-4\Omega^2\gamma^{AB}\zeta_B=O(1)$. 
Integrating from $0$ to any $u^+\in[0,\delta]$, with $\beta^A=0$ at $u^+=0$, 
we obtain
\begin{eqnarray}
\beta^A= O(\delta).
\end{eqnarray}
Equations~\eqref{defv} and \eqref{BS} give ${\cal D}_a\Omega=O(1)$. 
Using $\rho_{+{\rm gw}}=\frac12\tilde\theta_{+ab}\tilde\theta_+^{~ab}$, the 
derivative bounds in Eq.~\eqref{flux-derivative-bounds}, together with 
Eqs.~\eqref{BS} and \eqref{RS}, then imply 
${\cal D}_a(\Omega^2\rho_{+{\rm gw}}),\,{\cal D}_a(\Omega^2R_{++})
=O(\delta^{-3/2})$. 
Therefore, one can see that the angular advection terms in 
Eqs.~\eqref{flux-remainder-bounds}--\eqref{flux-remainder-bounds-matter} are indeed of 
order $\delta^{-1/2}$.

We next verify the gravitational-wave source bound in 
Eq.~\eqref{flux-source-bounds}. 
Under Eqs.~\eqref{BS} and \eqref{angular-derivative-counting} alone, 
$S_{+ab}\tilde\theta_+^{~ab}$ would contain the contribution 
$\tilde\theta_+^{~ab}{\cal D}_av_{+b}=O(\delta^{-1})$, of the same order as the 
transport term in Eq.~\eqref{terhogw}. 
Since only the symmetric traceless part of ${\cal D}_av_{+b}$ contributes to 
this contraction, the bound in Eq.~\eqref{angularBS} gives 
$\tilde\theta_+^{~ab}{\cal D}_av_{+b}=O(\delta^{-1/2})$. 
The remaining terms in Eq.~\eqref{S+ab} are at most of this order by 
Eqs.~\eqref{BS} and \eqref{RS}, and the bounds in Eqs.~\eqref{BS}--\eqref{flux-derivative-bounds} give $S_+=O(\delta^{-1/2})$ for the matter source in Eq.~\eqref{S+}.

\subsection{Consistency with the evolution equations}
\label{sec:evolution-consistency}

We now confirm that the assumed orders are consistent with the evolution equations. 
Unlike the bounds in Sec.~\ref{ssSec:OA}, the orders of the null derivatives are not assumed: 
they are read off from the evolution equations, and we check that integrating them reproduces the assumed orders. 
In Sec.~\ref{ssSec:CD}, we specify how derivatives along $K_\pm$ are to be counted in this check, 
and in Sec.~\ref{ssSec:OCEE} we examine the individual evolution equations. 
The check concerns the powers of $\delta$; we do not control the constants in the estimates, 
which would be needed to close a bootstrap argument.

\subsubsection{Counting of derivatives along \texorpdfstring{$K_\pm$}{K±}}
\label{ssSec:CD}

Suppose that $\psi$ is a variable of order $\delta^s$,
{\it i.e.} $\psi = O(\delta^s)$.
For the quantities and null directions considered below,
we check whether the evolution equations yield
\begin{align}
\mbox \pounds_{K_+} \psi= O(\delta^{s-1}),
\qquad
\mbox \pounds_{K_-} \psi= O(\delta^s).
\label{derivative-counting}
\end{align}
For surface tensors, we use components in the original double-null coordinates. 
Their orders agree with those in local orthonormal frames by 
Eq.~\eqref{angular-metric-estimate}. 
Since $\beta^A=O(\delta)$ and $\partial_A\beta^B=O(\delta^{1/2})$, the 
difference between $\Omega \mbox \pounds_{K_-}\psi$ and $\partial_-\psi$ is 
$O(\delta^{s+1/2})$ and does not change the orders in 
Eq.~\eqref{derivative-counting}. 
Since $\Omega=O(1)$, integrating the first bound in 
Eq.~\eqref{derivative-counting} over a $u^+$ interval of length at most 
$\delta$ gives a change of $O(\delta^s)$. 
Similarly, the second bound gives a change of $O(\delta^s)$ over the fixed 
$u^-$ interval $[-r_0,-r_f]$. 
Therefore, obtaining the relevant bounds in Eq.~\eqref{derivative-counting} 
from the evolution equations confirms that the assumed order 
$\psi=O(\delta^s)$ is self-consistent at the level of powers of $\delta$, 
provided that the initial data have this order.

If $\psi$ satisfies Eq.~\eqref{derivative-counting}, then ${\cal D}\psi$ 
satisfies the same bounds with $s$ replaced by $s-1/2$, provided that the 
angular derivative counting in Eq.~\eqref{angular-derivative-counting} applies 
to the right-hand sides of the evolution equations and their required angular 
derivatives. 
A derivation is given in Appendix~\ref{app:angular-commutators}.

We note that ${\cal R}=O(\delta^{1/2})$ follows from Eq.~\eqref{RS}, so that 
the assumed bound in Eq.~\eqref{ricci-trace-derivative-bound} agrees with 
Eq.~\eqref{derivative-counting} for $s=1/2$.

In what follows, we examine the evolution equations
in the null directions specified below for each quantity.

\subsubsection{Order consistency of evolution equations}
\label{ssSec:OCEE}

We examine the order consistency associated with the $K_+$ evolution of
$\Omega-1$, $\theta_+$, $\tilde\theta_{-ab}$,
$\omega_-$, $\zeta_a$, and ${}^{(2)}R$,
and the $K_-$ evolution of
$\Omega-1$, $\Delta\theta_-$, $\tilde\theta_{+ab}$,
$\omega_+$, and $R_{++}$.
The variables $v_{\pm a}$ are determined by $\Omega$ and 
$\zeta_a$ through Eq.~\eqref{defv}, and therefore need no separate check.
Note that the stronger angular derivative bounds in Eqs.~\eqref{angularBS} and 
\eqref{angular-laplacian-bound} are not needed for these checks themselves.

We first estimate the intrinsic curvature of the
two-surfaces. The bounds in Eq.~\eqref{BS} and the angular
derivative counting in Eq.~\eqref{angular-derivative-counting}
give $O(\delta^{-3/2})$ for the angular derivative terms
in Eq.~\eqref{2Ricci}.
Multiplying this equation by $\Omega$ gives a transport
equation in $u^+$ with a uniformly bounded coefficient
$\Omega\theta_+$.
Integrating over $0\le u^+\le\delta$, with the Minkowski
boundary value ${}^{(2)}R=O(1)$ at $u^+=0$, yields
\begin{eqnarray}
{}^{(2)}R = O(\delta^{-1/2}) .
\label{2RS}
\end{eqnarray}
Equation~\eqref{2Ricci} then gives
$\mbox \pounds_{K_+}{}^{(2)}R=O(\delta^{-3/2})$,
in agreement with Eq.~\eqref{derivative-counting}
for $s=-1/2$.

For $\Omega-1$, Eq.~\eqref{defomega} gives 
$\mbox \pounds_{K_\pm}(\Omega-1)=2\Omega\omega_\pm$, and the bounds in 
Eq.~\eqref{BS} imply $\mbox \pounds_{K_+}(\Omega-1)=O(\delta^{-1/2})$ and 
$\mbox \pounds_{K_-}(\Omega-1)=O(\delta^{1/2})$.

For $\theta_+$, the Raychaudhuri equation
\eqref{Ray}, together with Eqs.~\eqref{BS} and \eqref{RS},
gives $\mbox \pounds_{K_+}\theta_+=O(\delta^{-1})$.
The squared shear and $R_{++}$ contribute at this order,
while the remaining terms are smaller.

For $R_{++}$, Eq.~\eqref{bianchi}, together with
Eq.~\eqref{defomega}, the bounds in
Eqs.~\eqref{BS} and \eqref{RS}, and
$S_+=O(\delta^{-1/2})$ obtained in the previous
subsection, gives
$\mbox \pounds_{K_-}R_{++}=O(\delta^{-1})$.
This agrees with Eq.~\eqref{derivative-counting}
for $s=-1$.

For the shears, we use the traceless parts of
Eq.~\eqref{ggLt}.
The bounds in Eqs.~\eqref{BS} and \eqref{RS},
together with the angular derivative counting in
Eq.~\eqref{angular-derivative-counting}, give
$\mbox \pounds_{K_+}\tilde\theta_{-ab}=O(\delta^{-1/2})$
and
$\mbox \pounds_{K_-}\tilde\theta_{+ab}=O(\delta^{-1/2})$.

For $\omega_\pm$, we use
Eqs.~\eqref{evomega-} and \eqref{evomega+}
in Appendix~\ref{app:connection}.
With the curvature estimate \eqref{2RS}
and the assumed bounds, these equations give
$\mbox \pounds_{K_+}\omega_-=O(\delta^{-1/2})$
and
$\mbox \pounds_{K_-}\omega_+=O(\delta^{-1/2})$.
All these estimates agree with
Eq.~\eqref{derivative-counting}.

Next, for $\Delta \theta_-$, we use Eq.~\eqref{Ray-} with the contribution of the time evolution of $\theta_{0-}$ subtracted, which is 
\begin{eqnarray}
\mbox \pounds_{K_-} \Delta \theta_- &=& 2\omega_- \Delta \theta_- -\theta_{0-}\Delta \theta_- -\frac12 (\Delta \theta_-)^2
-\tilde \theta_{-ab} \tilde \theta_-^{~ab}-R_{--}
\nonumber \\
&&\quad +2\omega_-\theta_{0-}+\frac12(\Omega^{-1}-1)\theta_{0-}^2.
\end{eqnarray}
Using $\theta_{0-}=O(1)$ from
Eq.~\eqref{Minkexpansions} and
$\Omega^{-1}-1=O(\delta^{1/2})$,
the bounds in Eqs.~\eqref{BS} and \eqref{RS}
show that every term on the right-hand side is
$O(\delta^{1/2})$ or smaller.
Hence,
$\mbox \pounds_{K_-}\Delta\theta_-=O(\delta^{1/2})$,
in agreement with the $K_-$ bound in
Eq.~\eqref{derivative-counting} for $s=1/2$.

For $\zeta_a$, Eq.~\eqref{null-AB}
gives the evolution equation along $K_+$:
\begin{eqnarray}
\mbox \pounds_{K_+} \zeta_a &=& - R_{bc} {\gamma_a}^b K_+^c  -2 {\cal D}_a \omega_+  - {\cal D}_a \theta_+ + {\cal D}_b {\theta_{+a}}^b 
\nonumber \\
&&\qquad
-  \theta_+ \zeta_a  
- 2 \omega_+ {\cal D}_a\ln \Omega
+ {\theta_{+a}}^b{\cal D}_b\ln \Omega   
-2 K_{+a} \zeta_b \zeta^b.
\end{eqnarray}
The bounds in Eqs.~\eqref{BS} and \eqref{RS},
together with the angular derivative counting in
Eq.~\eqref{angular-derivative-counting}, give
${\cal D}_a\omega_+=O(\delta^{-1})$ and
${\cal D}_b{\theta_{+a}}^b=O(\delta^{-1})$.
All remaining terms are $O(\delta^{-1/2})$ or smaller.
Hence,
$\mbox \pounds_{K_+}\zeta_a=O(\delta^{-1})$,
in agreement with Eq.~\eqref{derivative-counting}
for $s=0$.

The stronger angular derivative bounds in Eqs.~\eqref{angularBS} and 
\eqref{angular-laplacian-bound} are not covered by the general commutator 
estimates in Appendix~\ref{app:angular-commutators}, and require a separate 
consistency check. 
They are examined using their evolution equations along $K_+$ and $K_-$ in 
Appendices~\ref{Sec+Dv+} and \ref{SecD2t}, respectively. 
The spacetime curvature estimates used in these checks are collected in 
Appendix~\ref{app:curvature}.

\section{Summary and discussion}
\label{sec:summary}

We have discussed the formation of trapped surfaces caused by a short pulse of 
incoming gravitational radiation and matter in a double-null setting. 
Under specified order estimates throughout the region of interest, we have 
derived a sufficient condition for trapped surface formation in terms of the 
initial incoming fluxes. 
The derivation combines the focusing of outgoing null geodesics through the 
Raychaudhuri equation with approximately conservative inward transport of the 
gravitational-wave and matter fluxes, in which the two enter on an equal 
footing. 
For the matter contribution, the contracted Bianchi identity identifies the 
terms that modify conservative transport, and we give conditions on the 
geometry, the Ricci tensor, and their derivatives under which these terms are 
subleading.

For the Minkowski initial region considered here, the sufficient condition is 
given by Eq.~\eqref{initialcriterion}. 
Here $r_0$ is the radius of the initial two-sphere on which the incoming flux 
is prescribed, and $r_f$ is the parameter specifying the final two-surface, kept fixed as 
the pulse width $\delta$ is made small. 
For fixed $0<r_f<r_0$, this condition must hold at every angle with a margin 
exceeding a uniform bound on the $O(\delta^{1/2})$ correction. 
Together with the assumed estimate for the incoming expansion, it ensures that 
the final two-surface is trapped for sufficiently small $\delta$. 
The physical mechanism is the amplification of the incoming fluxes as the 
cross-sectional area of the incoming null congruence shrinks. 
At leading order, the fluxes increase by a factor $(r_0/r_f)^2$, whereas the 
focusing threshold increases only by a factor $r_0/r_f$. 
The formation condition is therefore compatible with initially untrapped
sections when the integrated initial flux satisfies the upper bound
discussed after Eq.~\eqref{initialsurface}, and its combined focusing
term obeys, for example, Eq.~\eqref{combined-focus-initial}.

In Sec.~\ref{sec:order-consistency}, we introduce a more detailed hierarchy of 
orders for the geometric quantities, the Ricci components, and their 
derivatives, from which the bounds on the angular advection and source terms 
used in the formation argument follow. 
We also read off the orders of the derivatives from the
evolution equations considered and check their agreement with the assumed orders, 
including the stronger angular derivative bounds examined in 
Appendices~\ref{Sec+Dv+} and \ref{SecD2t}. 
Appendix~\ref{app:angular-commutators} gives a general commutator argument for 
extending these estimates to angular derivatives under the assumed angular 
derivative counting. 
These checks concern only the powers of $\delta$. 
That is, we assume the existence of a regular spacetime with a regular 
double-null foliation, and the validity of the required bounds with constants 
independent of $\delta$, rather than establishing them from initial data.

In vacuum, the resulting criterion agrees at leading order with the 
established short-pulse formation 
criterion~\cite{Christodoulou2009,KlainermanRodnianski2012,BiniEsposito2017}. 
The additive contributions of gravitational radiation and matter also appear 
in results for specific Einstein--matter 
systems~\cite{Yu:2011qy,AnAthanasiou2020,ZhaoHilditchValienteKroon2023,AthanasiouMondalYau2023,Shen:2025ztd}. 
In each case, however, the analysis relies on the field equations of the 
specific matter model. 
Here we instead formulate the matter contribution through conditions on the 
Ricci tensor and its derivatives, so that the formation argument can be 
developed without specifying a particular matter model. 
These conditions therefore provide guidance on the matter models to which the mechanism 
may apply and on the initial data that should be prescribed for them.

Since the conditions constrain only the Ricci components and their 
derivatives, they can be examined for a given model using the Einstein 
equation before attempting the full estimates. They may therefore serve as a first 
step towards studying matter systems that have not yet been treated. 
One such case is a nonzero cosmological constant, which contributes to the 
Ricci tensor through the metric, but which also requires separating its 
background contribution in the Ricci estimates and replacing the Minkowski 
diamond by the corresponding constant-curvature background. 
The transport estimates and formation criterion would then need to be 
reexamined. 
For a positive cosmological constant, this may allow the interplay between 
cosmic expansion and gravitational collapse to be analyzed along the lines of 
Sec.~\ref{TSF}. 
We leave these extensions for future work.

\acknowledgments
We thank the participants in the Shirataka Workshop on Astrophysics, held in Shirataka, Yamagata Prefecture, the hometown of the late Professor 
Humitaka Sato, for their valuable comments. 
This work was supported by Grants-in-Aid for Scientific Research from the Ministry of Education, 
Science, Sports and Culture of Japan (No. 21H05189).
T.~S.~and K.~I.~were supported by Grants-in-Aid for Scientific Research from the Ministry of Education, 
Science, Sports and Culture of Japan (No. 21H05182). T.~S.~was also supported by JSPS Grants-in-Aid for Scientific Research (C) (JP25K07320), 
the Fund for the Promotion of Joint International
Research (JP23KK0048) and Grant-in-Aid for Scientific Research (A) (JP24H00183). 
K. I. was also supported by JSPS Grant-in-Aid for Scientific Research (C) (JP24K07046). D.~Y.~was also supported by 
JSPS Grant-in-Aid for Scientific Research (C) (JP26K07084). 

\appendix

\section{Evolution of \texorpdfstring{$\omega_\pm$}{the null connection coefficients}}\label{app:connection}

The evolution equations for $\omega_\pm$
can be obtained from
$R_{+-+-}:=R_{abcd}K_+^a K_-^b K_+^c K_-^d$,
which satisfies
\begin{eqnarray}
R_{+-+-} &=& -K_+^a K_-^b K_-^c (\nabla_a \nabla_b-\nabla_b \nabla_a) K_{+c} \nonumber \\
&=& - 4 \left( \mbox \pounds_{K_+} \omega_- + \mbox \pounds_{K_-} \omega_+ \right)- 16 \omega_+ \omega_- +4v_{+a} v_-^a - 8\zeta_a\zeta^a .
\end{eqnarray}
Using the relation
\begin{eqnarray}
 \mbox \pounds_{K_-} \omega_+= \mbox \pounds_{K_+} \omega_- + 2 \zeta^a \partial_a \ln \Omega,
\end{eqnarray}
which follows from the definition~\eqref{defomega} and 
\begin{eqnarray}
[K_+,K_-]^a\nabla_a\ln\Omega = -4\zeta^a\partial_a\ln\Omega,
\end{eqnarray}
we can rewrite the above equation as
\begin{eqnarray}
R_{+-+-} &=& - 8 \mbox \pounds_{K_+} \omega_-  - 8 \zeta^a \partial_a \ln \Omega - 16 \omega_+ \omega_- + 4v_{+a} v_-^{~a} - 8\zeta_a\zeta^a 
\nonumber\\
&=& - 8 \mbox \pounds_{K_-} \omega_+  + 8 \zeta^a \partial_a \ln \Omega - 16 \omega_+ \omega_- + 4v_{+a} v_-^{~a} - 8\zeta_a\zeta^a.
\end{eqnarray}
Moreover, $R_{+-+-}$ is expressed in terms of the Ricci tensor and other geometric quantities as
\begin{eqnarray}
R_{+-+-} = 2R_{+-}+ 2{{\cal R}} - 2 {}^{(2)}R + 2 \theta_{+ab}\theta_-^{~ab} - 2 \theta_+ \theta_-.
\end{eqnarray}
Therefore, $\mbox \pounds_{K_+} \omega_-$ and $\mbox \pounds_{K_-} \omega_+$ become
\begin{eqnarray}
\mbox \pounds_{K_+} \omega_- &=& -\frac14 R_{+-} - \frac14{{\cal R}} + \frac14 {}^{(2)}R - \frac14 \theta_{+ab}\theta_-^{~ab} + \frac14 \theta_+ \theta_-
\nonumber\\ && \quad
-  \zeta^a \partial_a \ln \Omega - 2 \omega_+ \omega_- + \frac12 v_{+a} v_-^{~a} - \zeta_a\zeta^a,
\label{evomega-} 
\\
\mbox \pounds_{K_-} \omega_+ &=&  -\frac14 R_{+-} - \frac14{{\cal R}} + \frac14 {}^{(2)}R - \frac14 \theta_{+ab}\theta_-^{~ab} + \frac14 \theta_+ \theta_-
\nonumber\\ && \quad
+  \zeta^a \partial_a \ln \Omega - 2 \omega_+ \omega_- + \frac12 v_{+a} v_-^{~a} - \zeta_a\zeta^a.
\label{evomega+}
\end{eqnarray}

\section{Commutators with angular derivatives}
\label{app:angular-commutators}

We derive the commutator estimates used in
Sec.~\ref{sec:evolution-consistency}.
For simplicity, we present the derivation for a covector
$V_a$ tangent to the two-surfaces
of constant $u^+$ and $u^-$.
The derivation for higher-rank covariant tensors is analogous.
We define the projected Lie derivative by
\begin{equation}
\widehat{\mbox{\pounds}}_X V_a
:=\gamma_a{}^b\mbox{\pounds}_X V_b.
\end{equation}
For higher-rank covariant tensors, all indices are
projected onto the two-surfaces.

We take $X$ to have a flow mapping the two-surfaces to one another, 
so that its projected Lie derivative commutes with ${\cal D}_a$ on scalars.  
On covectors, we write
\begin{equation}
[\widehat{\mbox{\pounds}}_X,{\cal D}_a]V_b
=-A_{ab}{}^cV_c.
\end{equation}
The torsion-free property of ${\cal D}_a$ implies
$A_{ab}{}^c=A_{ba}{}^c$.
Writing $h_{ab}:=\widehat{\mbox{\pounds}}_X\gamma_{ab}$, the metric compatibility gives
\begin{equation}
{\cal D}_a h_{bc}
=A_{ab}{}^d\gamma_{dc}+A_{ac}{}^d\gamma_{bd}.
\end{equation}
Permuting the indices and using the symmetry of
$A_{ab}{}^c$, we obtain
\begin{equation}
A_{ab}{}^c
=\frac12\gamma^{cd}
\left(
{\cal D}_a h_{bd}
+{\cal D}_b h_{ad}
-{\cal D}_d h_{ab}
\right).
\end{equation}

Applying this to $X=\Omega K_\pm$, for which $h_{ab}=2\Omega\theta_{\pm ab}$, 
and using $\hat{\mbox \pounds}_{K_\pm}=\Omega^{-1}\hat{\mbox \pounds}_{\Omega K_\pm}$ on 
tangential covariant tensors, we obtain
\begin{equation}
[\widehat{\mbox{\pounds}}_{K_\pm},{\cal D}_a]V_b
=({\cal D}_a\ln\Omega)
 \widehat{\mbox{\pounds}}_{K_\pm}V_b
-C_{\pm ab}{}^cV_c,
\label{angular-lie-commutator}
\end{equation}
where
\begin{equation}
C_{\pm ab}{}^c
:=\Omega^{-1}\left[
{\cal D}_a(\Omega\theta_{\pm b}{}^c)
+{\cal D}_b(\Omega\theta_{\pm a}{}^c)
-{\cal D}^c(\Omega\theta_{\pm ab})
\right].
\end{equation}
For a scalar, the term involving $C_{\pm ab}{}^c$
is absent. For a higher-rank covariant tensor,
each covariant index contributes a corresponding term.

Equations~\eqref{defv},~\eqref{BS} and
\eqref{angular-derivative-counting} give
${\cal D}_a\ln\Omega=O(1)$,
$C_{+ ab}{}^c=O(\delta^{-1})$, and
$C_{- ab}{}^c=O(\delta^{-1/2})$.
Thus, if $\psi=O(\delta^s)$ satisfies
Eq.~\eqref{derivative-counting}, then
\begin{equation}
[\widehat{\mbox{\pounds}}_{K_+},{\cal D}]\psi
=O(\delta^{s-1}),
\qquad
[\widehat{\mbox{\pounds}}_{K_-},{\cal D}]\psi
=O(\delta^{s-1/2}).
\label{commutator-estimates}
\end{equation}
The evolution of ${\cal D}\psi$ is given by
\begin{eqnarray}
\hat{\mbox \pounds}_{K_\pm}({\cal D}_a\psi) 
= {\cal D}_a(\hat{\mbox \pounds}_{K_\pm}\psi) + [\hat{\mbox \pounds}_{K_\pm}, {\cal D}_a]\psi .
\label{commutator-split}
\end{eqnarray}
Applying the angular derivative counting to the right-hand sides of the 
evolution equations and their required angular derivatives gives 
${\cal D}(\hat{\mbox \pounds}_{K_+}\psi) = O(\delta^{s-3/2})$ and 
${\cal D}(\hat{\mbox \pounds}_{K_-}\psi) = O(\delta^{s-1/2})$. 
Together with Eq.~\eqref{commutator-estimates},  neither term on the 
right-hand side of Eq.~\eqref{commutator-split} exceeds these orders, so that 
$\hat{\mbox \pounds}_{K_+}({\cal D}\psi) = O(\delta^{s-3/2})$ and 
$\hat{\mbox \pounds}_{K_-}({\cal D}\psi) = O(\delta^{s-1/2})$.

Finally, we account for the normal components of
the unprojected Lie derivatives.
The tangential part of $[K_+,K_-]^a$ is $-4\zeta^a$.
Using $V_aK_\pm^a=0$, we obtain
\begin{equation}
\mbox{\pounds}_{K_\pm}V_a
=\widehat{\mbox{\pounds}}_{K_\pm}V_a
\mp2K_{\pm a}\zeta^bV_b.
\end{equation}
For higher-rank covariant tensors, there is one such
normal term for each index.
Since $\zeta_a=O(1)$, the normal components of
$\mbox{\pounds}_{K_\pm}({\cal D}\psi)$ are
$O(\delta^{s-1/2})$.
Consequently,
\begin{equation}
\mbox{\pounds}_{K_+}({\cal D}\psi)
=O(\delta^{s-3/2}),
\qquad
\mbox{\pounds}_{K_-}({\cal D}\psi)
=O(\delta^{s-1/2}),
\end{equation}
which are the bounds in Eq.~\eqref{derivative-counting}
with $s$ replaced by $s-1/2$.
This argument assumes the angular derivative bounds;
it does not establish them from initial data.

\section{Estimate of the Riemann tensor}\label{app:curvature}

In the analysis presented in Appendices~\ref{Sec+Dv+} and \ref{SecD2t}, an estimate of the Riemann tensor is required. 
We derive this estimate here.

We denote the projected Riemann and Ricci tensors, some of which have already been defined, by
\begin{eqnarray}
&&{\cal R}_{abcd} := R_{efgh}  {\gamma_a}^e {\gamma_b}^f {\gamma_c}^g {\gamma_d}^h, \qquad
{\cal R}_{\pm abc} := R_{efgh} K_\pm^e {\gamma_a}^f {\gamma_b}^g {\gamma_c}^h, 
\nonumber \\
&& 
{\cal R}_{\pm a\pm b} := R_{efgh} K_\pm^e {\gamma_a}^f K_\pm^g {\gamma_b}^h, \qquad
{\cal R}_{+ a- b} := R_{efgh} K_+^e {\gamma_a}^f K_-^g {\gamma_b}^h, 
\nonumber \\
&&
{\cal R}_{+ - a b} := R_{efgh} K_+^e K_-^f {\gamma_a}^g  {\gamma_b}^h, \qquad
{\cal R}_{\pm \mp \pm a} := R_{efgh} K_\pm^e K_\mp^f K_\pm^g {\gamma_a}^h, 
\nonumber \\
&& 
{\cal R}_{ab} := R_{cd} {\gamma_a}^c {\gamma_b}^d, \qquad
{\cal R}_{\pm a} := R_{cd} K_\pm^c {\gamma_a}^d, 
\end{eqnarray}
and other components by
\begin{eqnarray}
&&R_{+-+-} := R_{efgh}  K_+^e K_-^f K_+^g K_-^h, \qquad
R_{\pm\pm}:= R_{cd} K_\pm^c K_\pm^d, \qquad
R_{+-}:= R_{cd} K_+^c K_-^d.
\end{eqnarray}
The Ricci tensor is determined by the matter distribution through the Einstein equation. 
The components of the Riemann tensor other than $R_{+-+-}$ are
expressed in terms of $\omega_\pm$, $\zeta_a$, $v_{\pm a}$,
$\theta_{\pm}$, $\theta_{\pm ab}$, ${}^{(2)}R$,
the Ricci tensor, and derivatives of these quantities as
\begin{eqnarray}
&&{\cal R}_{abcd} =\frac12 {}^{(2)}R\left(\gamma_{ac}\gamma_{bd}-\gamma_{ad}\gamma_{bc}\right)+\frac{1}{2}
(\theta_{+ac}\theta_{-bd}+\theta_{-ac}\theta_{+bd}-\theta_{+bc}\theta_{-ad}-\theta_{-bc}\theta_{+ad}),\\
&&{\cal R}_{\pm abc} = -{\cal D}_b \theta_{\pm ac} + {\cal D}_c \theta_{\pm ab} \mp \theta_{\pm ac} \zeta_b \pm \theta_{\pm ab} \zeta_c,
\\
&& 
{\cal R}_{+ a- b} = -{\cal R}_{ab} + \frac12 {}^{(2)} R \gamma_{ab} - {\cal D}_a \zeta_b +  {\cal D}_b \zeta_a  - {\theta_{+a}}^c \theta_{-cb}  + \frac12 \left( \theta_+ \theta_{-ab} + \theta_- \theta_{+ab} \right)
\nonumber \\&& \qquad\quad=
-{\cal R}_{ab} + \frac12 {}^{(2)} R \gamma_{ab} - {\cal D}_a \zeta_b +  {\cal D}_b \zeta_a  + \frac14 \theta_+ \theta_- \gamma_{ab} - \tilde \theta_{+a}{}^c \tilde \theta_{-bc}
,
\\
&& 
{\cal R}_{\pm a \pm b} = -{\gamma_a}^c{\gamma_b}^d \mbox \pounds_{K_\pm} \theta_{\pm cd}+ \theta_{\pm ac} {\theta_{\pm b}}^c+ 2 \omega_\pm \theta_{\pm ab}
,
\\
&& 
{\cal R}_{+ - a b} =  2{\cal D}_b \zeta_a - 2{\cal D}_a \zeta_b+ \theta_{-ac}{\theta_{+b}}^c- \theta_{+ac}{\theta_{-b}}^c,
\\
&& 
{\cal R}_{\pm \mp \pm a} = 2 {\cal R}_{\pm a} +2{\cal D}_a \theta_{\pm} - 2{\cal D}_b {\theta_{\pm a}}^b \pm 2\theta_\pm \zeta_a \mp 2{\theta_{\pm a}}^b \zeta_b.
\end{eqnarray}
The order estimates of Eqs.~\eqref{BS}, \eqref{RS}, \eqref{angular-derivative-counting}
and \eqref{2RS} imply
the following bounds, subject to the additional
assumptions stated below for ${\cal R}_{\pm a\pm b}$
and ${\cal R}_{-+-a}$:
\begin{eqnarray}
&&{\cal R}_{abcd} = O(\delta^{-1/2}), \quad
{\cal R}_{+ abc} =O(\delta^{-1}),  \quad
{\cal R}_{- abc} =O(\delta^0),  
\nonumber \\
&&
{\cal R}_{+ a- b} = O(\delta^{-1/2}),\quad 
{\cal R}_{+ a+ b} = O(\delta^{-3/2}),\quad
{\cal R}_{- a- b} = O(\delta^{0}),
\nonumber \\
&&
{\cal R}_{+- a b} = O(\delta^{-1/2}),\quad
{\cal R}_{+-+ a} =O(\delta^{-1}), \quad
{\cal R}_{-+- a} =O(\delta^{0}).
\end{eqnarray}
The estimates for ${\cal R}_{+a+b}$ and
${\cal R}_{-a-b}$ require the additional derivative bounds
\begin{align}
{\gamma_a}^c{\gamma_b}^d
\mbox \pounds_{K_+}\tilde\theta_{+cd}
&=O(\delta^{-3/2}),\\
{\gamma_a}^c{\gamma_b}^d
\mbox \pounds_{K_-}\tilde\theta_{-cd}
&=O(1),
\end{align}
respectively.
These are not obtained from the evolution equations considered in 
Sec.~\ref{sec:evolution-consistency}: the $K_+$ evolution of 
$\tilde\theta_{+ab}$ and the $K_-$ evolution of $\tilde\theta_{-ab}$ do not 
appear there. 
The orders are chosen according to the derivative counting in 
Eq.~\eqref{derivative-counting}. 
The first bound coincides with it, while the second is weaker.
The estimate for ${\cal R}_{-+- a}$
also requires ${\cal R}_{- a}=O(\delta^{0})$,
which is not included in the assumptions for the Ricci tensor
in Eq.~\eqref{RS}.
Since these three curvature estimates are
not used in Appendices~\ref{Sec+Dv+} and \ref{SecD2t},
none of these additional assumptions is needed
for those analyses.

\section{Estimate of \texorpdfstring{$\tilde \theta_+^{~ab}{\cal D}_a v_{+b}$}{the shear contraction with the gradient of v}}\label{Sec+Dv+}

In this appendix, we check the consistency of the bound in
Eq.~\eqref{angularBS}
by verifying the $K_+$ bound in
Eq.~\eqref{derivative-counting} with $s=0$,
using Eq.~\eqref{angular-laplacian-bound},
while the consistency of the bound in
Eq.~\eqref{angular-laplacian-bound} is examined
in Appendix~\ref{SecD2t}.
Since only the symmetric traceless components of ${\cal D}_a v_{+b}$, namely $\left({\cal D}_a v_{+b}\right)^{(STL)}$, are involved in $\tilde \theta_+^{~ab}{\cal D}_a v_{+b}$, 
only a bound on these components is required,
as specified in Eq.~\eqref{angularBS}.
We will show that its Lie derivative along $K_+$,
{\it i.e.}
$\mbox \pounds_{K_+}\left({\cal D}_a v_{+b}\right)^{(STL)}$,
is of order $\delta^{-1}$.

The Leibniz rule for the Lie derivative gives
\begin{eqnarray}
\mbox \pounds_{K_+} \left({\cal D}_a v_{+b}\right)^{(STL)} &=& 
\frac12 \left(\mbox \pounds_{K_+} {\cal D}_c v_{+d}\right) \left({\gamma_a}^c{\gamma_b}^d+{\gamma_a}^d{\gamma_b}^c-\gamma_{ab}\gamma^{cd}\right)
\nonumber \\ &&\quad
+ \frac12 \left( {\cal D}_c v_{+d}\right) \mbox \pounds_{K_+} \left({\gamma_a}^c{\gamma_b}^d+{\gamma_a}^d{\gamma_b}^c-\gamma_{ab}\gamma^{cd}\right).
\end{eqnarray}
Since $\mbox \pounds_{K_+} \gamma_{ab}$ is of order
$\delta^{-1/2}$, with the leading contribution
coming from the shear $\tilde\theta_{+ab}$
whose bound is given in Eq.~\eqref{BS},
the second term on the right-hand side is of order
$\delta^{-1}$.
The first term on the right-hand side is
$\left({\gamma_a}^c{\gamma_b}^d
\mbox \pounds_{K_+}{\cal D}_c v_{+d}\right)^{(STL)}$.
Therefore, it remains to show that this quantity
is of order $\delta^{-1}$.

We have
\begin{eqnarray}
{\gamma_c}^e {\gamma_d}^f \mbox \pounds_{K_+} {\cal D}_e v_{+f}
&=&{\cal D}_c\left({\gamma_d}^fK_+^e \nabla_ev_{+f}\right) +\left({\cal D}_c \ln \Omega \right)\left({\gamma_d}^fK_+^e \nabla_ev_{+f}\right)
\nonumber \\ && \quad
+{\cal R}_{+cde}v_+^{~e}
- \theta_{+cd}v_{+e}v_-^{~e} + \theta_{+ce}v_+^{~e}v_{-d}+ {\theta_{+d}}^{e}{\cal D}_c v_{+e}.
\label{+Dv+}
\end{eqnarray}
On the right-hand side of Eq.~\eqref{+Dv+},
all terms except the first are at most of order $\delta^{-1}$.
Combining Eqs.~\eqref{defv} and \eqref{null-AB} gives
\begin{equation}
{\gamma_d}^f K_+^e \nabla_e v_{+f}
={\cal R}_{+d}
+{\cal D}_d\theta_+
-{\cal D}_e{\theta_{+d}}^e
+\theta_+\zeta_d
+{\theta_{+d}}^e\zeta_e.
\label{vplus-evolution}
\end{equation}
This gives
${\gamma_d}^fK_+^e\nabla_ev_{+f}=O(\delta^{-1})$.
In a naive estimate, the first term on the right-hand side
of Eq.~\eqref{+Dv+} therefore appears to be
of order $\delta^{-3/2}$.
As we now show, however, the angular derivative bounds
in Eqs.~\eqref{angular-derivative-counting}
and \eqref{angular-laplacian-bound}
and the symmetric traceless condition improve this to
$\delta^{-1}$. 
Taking an angular derivative of
Eq.~\eqref{vplus-evolution}, we obtain
\begin{eqnarray}
{\cal D}_c\left({\gamma_d}^fK_+^e \nabla_ev_{+f}\right)
= {\cal D}_c{\cal R}_{+d}+ {\cal D}_c{\cal D}_d \theta_+ - {\cal D}_c{\cal D}_e {\theta_{+d}}^e + {\cal D}_c\left(\theta_+ \zeta_d+{\theta_{+d}}^{e}\zeta_e\right).
\end{eqnarray}
All terms except the third on the right-hand side are at most of order $\delta^{-1}$. 
Therefore, if the contribution of ${\cal D}_c{\cal D}_e {\theta_{+d}}^e$ is at most of order $\delta^{-1}$, 
$\left({\cal D}_a v_{+b}\right)^{(STL)}$ can be of order  $\delta^{0}$. 
Note that, since we are considering the symmetric traceless part $\left({\cal D}_a v_{+b}\right)^{(STL)}$, 
only the symmetric traceless part of  ${\cal D}_c{\cal D}_e {\theta_{+d}}^e$, namely $\left({\cal D}_c{\cal D}_e {\theta_{+d}}^e\right)^{(STL)}$, is involved. 

To estimate
$\left({\cal D}_c{\cal D}_e{\theta_{+d}}^e\right)^{(STL)}$,
we use a two-dimensional identity for symmetric
traceless tensors.\footnote{
Define $w_b={\cal D}^c\tilde\theta_{+cb}$.
Since the two-surfaces are two-dimensional,
antisymmetrization over three tangential indices gives
\[
\gamma_{[c}{}^d\gamma_a{}^e\gamma_{g]}{}^f=0.
\]
Contracting this identity with
$\gamma_{eb}{\cal D}_d\tilde\theta_{+f}{}^g$
and using the symmetry
and tracelessness of $\tilde\theta_{+ab}$
gives
\[
{\cal D}_c\tilde\theta_{+ab}
-{\cal D}_a\tilde\theta_{+cb}
=\gamma_{cb}w_a-\gamma_{ab}w_c.
\]
Applying ${\cal D}^c$ and using
$[{\cal D}^c,{\cal D}_a]\tilde\theta_{+cb}
={}^{(2)}R\,\tilde\theta_{+ab}$
gives
$({\cal D}_a w_b)^{(STL)}
=\frac12{\cal D}^2\tilde\theta_{+ab}
-\frac12{}^{(2)}R\,\tilde\theta_{+ab}$.
Equation~\eqref{stf-divergence-identity} follows
by adding the trace contribution from
$\theta_{+ab}=\tilde\theta_{+ab}+\frac12\theta_+\gamma_{ab}$.}
Decomposing $\theta_{+ab}$ into its trace and shear gives

\begin{align}
\left({\cal D}_a{\cal D}_c{\theta_{+b}}^c\right)^{(STL)}
={}&\frac12{\cal D}^2\tilde\theta_{+ab}
-\frac12{}^{(2)}R\,\tilde\theta_{+ab}+\frac12
\left({\cal D}_a{\cal D}_b\theta_+\right)^{(STL)}.
\label{stf-divergence-identity}
\end{align}

The first term is $O(\delta^{-1})$ by
Eq.~\eqref{angular-laplacian-bound}.
The second term has the same order by
Eqs.~\eqref{BS} and \eqref{2RS}.
The last term is also $O(\delta^{-1})$ by
$\theta_+=O(1)$ and the angular derivative counting
in Eq.~\eqref{angular-derivative-counting}.
As a result, 
$\left({\cal D}_a {\cal D}_c \theta_{+b}{}^c\right)^{(STL)}$ is of order $\delta^{-1}$, which yields 
\begin{eqnarray}
\mbox \pounds_{K_+} \left({\cal D}_a v_{+b}\right)^{(STL)}=O(\delta^{-1}).
\end{eqnarray}

\section{Estimate of \texorpdfstring{${\cal D}^2 \tilde \theta_{+ab}$}{the angular Laplacian of the shear}}\label{SecD2t}

In this appendix, we check the consistency of the bound
${\cal D}^2 \theta_{+ab}=O(\delta^{-1})$, or equivalently ${\cal D}^2 \tilde \theta_{+ab}=O(\delta^{-1})$,
in Eq.~\eqref{angular-laplacian-bound}
by verifying the $K_-$ bound in
Eq.~\eqref{derivative-counting} with $s=-1$.
Its Lie derivative along $K_-$ is
\begin{eqnarray}
\mbox \pounds_{K_-} {\cal D}^2 \theta_{+ab}& = & K_-^c \nabla_c{\cal D}^2 \theta_{+ab}
+{\cal D}^2 \theta_{+cb}\nabla_aK^c_-+{\cal D}^2 \theta_{+ac}\nabla_b K_-^c
\nonumber \\
& = &  K_-^c \nabla_c{\cal D}^2 \theta_{+ab}+{\cal D}^2 \theta_{+cb}(\theta_{-a}{}^{c}+K_{-a}v_-^c)
+{\cal D}^2 \theta_{+ca}(\theta_{-b}{}^{c}+K_{-b}v_-^c).
\label{-DDtheta+}
\end{eqnarray}
Note that only the traceless part of the above equation
needs to be analyzed
for the additional bound in
Eq.~\eqref{angular-laplacian-bound}, since
${\cal D}^2\theta_+=O(\delta^{-1})$ already follows from
$\theta_+=O(1)$ and the angular derivative counting
in Eq.~\eqref{angular-derivative-counting}.
The first term on the right-hand side of
Eq.~\eqref{-DDtheta+} is computed as\footnote{
To derive Eqs.~\eqref{KnablaD2theta} and
\eqref{incoming-angular-commutator}, we express
${\cal D}$ as the fully projected spacetime
covariant derivative and differentiate the
projectors using
\begin{align*}
K_-^c\nabla_c\gamma_a{}^b
&=-K_{-a}v_+^b-v_{+a}K_-^b,\\
\gamma_e{}^f\nabla_fK_-^g
&=\theta_{-e}{}^g+\zeta_eK_-^g.
\end{align*}
We then commute the spacetime covariant
derivatives and differentiate the orthogonality
relations between the tangential tensors
and $K_\pm^a$.
}
\begin{eqnarray}
K_-^c \nabla_c{\cal D}^2 \theta_{+ab}
={\cal D}^e
(K_-^g {\gamma_e}^h
{\gamma_a}^i {\gamma_b}^j
\nabla_g {\cal D}_h \theta_{+ij})
+F_{ab}^{\delta^{-1}},\label{KnablaD2theta}
\end{eqnarray}
where
\begin{eqnarray}
F^{\delta^{-1}}_{ab}& := & -\tilde \theta_-^{~cd}{\cal D}_c {\cal D}_d \theta_{+ab}-\frac{1}{2}\theta_- {\cal D}^2 \theta_{+ab}
+{\gamma_a}^e {\gamma_b}^f \left({\cal D}^c \ln \Omega \right) K_-^d \nabla_d {\cal D}_c \theta_{+ef} \nonumber \\
& & +v_+^{~c} \theta_{-c}{}^{d}{\cal D}_d \theta_{+ab}
+v_{+a}\theta_-^{~cd}{\cal D}_c \theta_{+bd}+v_{+b}\theta_-^{~cd}{\cal D}_c \theta_{+ad} 
- v_+^{~e} \theta_- {\cal D}_e \theta_{+ab}
\nonumber \\
& & 
- v_+^{~c} \theta_{-a}{}^d {\cal D}_{d} \theta_{+cb}
- v_+^{~c} \theta_{-b}{}^d {\cal D}_{d} \theta_{+ca}
-{\cal R}_{-d}^{~~~cd}{\cal D}_c \theta_{+ab}+{\cal R}_{-da}^{~~~~c}{\cal D}^d \theta_{+cb}+{\cal R}_{-db}^{~~~~c}{\cal D}^d \theta_{+ca} \nonumber \\
& & -K_{-a}v_+^{~c}{\cal D}^2 \theta_{+cb}-K_{-b}v_+^{~c}{\cal D}^2 \theta_{+ac}.
\end{eqnarray}
Under the bounds in Eqs.~\eqref{BS}--\eqref{angular-laplacian-bound} and \eqref{2RS},
all terms in $F^{\delta^{-1}}_{ab}$ except the third
are directly estimated as $O(\delta^{-1})$.
To estimate the mixed derivative in the third term,
we use the following commutation relation:
\begin{eqnarray}
K_-^g {\gamma_e}^h {\gamma_c}^i {\gamma_d}^j  \nabla_g {\cal D}_h \theta_{+ij}
& = & {\cal D}_e (K_-^g {\gamma_c}^i {\gamma_d}^j \nabla_g \theta_{+ij})-\frac{1}{2}\theta_- {\cal D}_e \theta_{+cd} +G^{\delta^{-1/2}}_{ecd},
\label{incoming-angular-commutator}
\end{eqnarray}
where 
\begin{eqnarray}
G^{\delta^{-1/2}}_{ecd} &:= & 
-\tilde \theta_{-e}{}^{g}{\cal D}_g \theta_{+cd}+{\cal R}_{-ec}^{~~~~g}\theta_{+gd}+{\cal R}_{-ed}^{~~~~g}\theta_{+gc} \nonumber \\
& & -\zeta_e {\gamma_c}^i {\gamma_d}^j K_-^g \nabla_g \theta_{+ij}-\theta_{-ec}\theta_{+dg}v_+^{~g}-\theta_{-ed}\theta_{+cg}v_+^{~g} \nonumber \\
& & -v_{+e}{\gamma_c}^i {\gamma_d}^j K_-^g \nabla_g \theta_{+ij}+v_{+c}\theta_{-e}{}^{g} \theta_{+gd}+v_{+d}\theta_{-e}{}^{g} \theta_{+gc}.
\end{eqnarray}
The bounds give
$G^{\delta^{-1/2}}_{ecd}=O(\delta^{-1/2})$.
Together with Eq.~\eqref{ggLt} and the angular derivative
counting in Eq.~\eqref{angular-derivative-counting},
Eq.~\eqref{incoming-angular-commutator} then gives
$O(\delta^{-1})$ for the mixed derivative on its
left-hand side.
Since ${\cal D}_a\ln\Omega=O(1)$, the third term in
$F^{\delta^{-1}}_{ab}$ is also $O(\delta^{-1})$,
completing the estimate
$F^{\delta^{-1}}_{ab}=O(\delta^{-1})$.

We now turn to the first term on the right-hand side
of Eq.~\eqref{KnablaD2theta}.
A naive estimate would assign this term
an order $O(\delta^{-3/2})$,
but, as we now show, its traceless part is
$O(\delta^{-1})$.
Substituting Eq.~\eqref{incoming-angular-commutator}
into Eq.~\eqref{KnablaD2theta} and using
Eq.~\eqref{-DDtheta+}, we obtain
\begin{eqnarray}
\mbox \pounds_{K_-}{\cal D}^2 \theta_{+ab}  = {\cal D}^2(K_-^e {\gamma_a}^c {\gamma_b}^d \nabla_e \theta_{+cd})
-\frac{1}{2}{\cal D}^c (\theta_- {\cal D}_c \theta_{+ab})+O(\delta^{-1}). 
\end{eqnarray}
Since only its traceless part is relevant to the time evolution of the order assumption, we therefore consider its traceless part.
From Eq.~\eqref{ggLt}, the traceless part of $K_-^e {\gamma_a}^c {\gamma_b}^d \nabla_e \theta_{+cd}$ in the first term 
is $-(1/2)\theta_- \tilde \theta_{+ab}+O(\delta^0)$. It follows that\footnote{
Here the $O(\delta^0)$ traceless remainder
obeys the angular derivative counting
in Eq.~\eqref{angular-derivative-counting},
so that two angular derivatives give $O(\delta^{-1})$.
In particular, this uses the assumption stated below
Eq.~\eqref{angular-laplacian-bound} for
the angular derivatives of the improved
$({\cal D}_av_{+b})^{(STL)}$ term.
} 
\begin{eqnarray}
\mbox \pounds_{K_-} ({\cal D}^2 \theta_{+ab})^{(STL)}= 
-\theta_-  \left({\cal D}^2 \tilde \theta_{+ab}\right)
-\frac32 \left({\cal D}^c \theta_-\right) \left({\cal D}_c \tilde \theta_{+ab}\right)
-\frac{1}{2}\left({\cal D}^2 \theta_- \right) \tilde \theta_{+ab}+O(\delta^{-1}).
\end{eqnarray}
Note that, since $\theta_{0-}$
is extended to ${\cal M}_1$ using
Eq.~\eqref{Minkexpansions} and depends only on $u^\pm$,
its angular derivative vanishes. Thus, we have
\begin{eqnarray}
{\cal D}_a \theta_-={\cal D}_a \Delta\theta_-=O(\delta^0),\qquad {\cal D}^2 \theta_-={\cal D}^2 \Delta\theta_-=O(\delta^{-1/2}).
\end{eqnarray}
Together with the angular derivative bound
${\cal D}^2 \tilde \theta_{+ab}=O(\delta^{-1})$,
we find
$\mbox \pounds_{K_-} ({\cal D}^2 \theta_{+ab})^{(STL)}
=O(\delta^{-1})$.
Therefore, the angular derivative bound
in Eq.~\eqref{angular-laplacian-bound}
is consistent with its evolution equation.

\end{document}